\documentclass[acmsmall]{acmart}
\usepackage{graphicx}
\usepackage[skip=0pt]{caption}
\usepackage{subcaption}
\usepackage[most]{tcolorbox}
\usepackage{xcolor}
\usepackage{xspace}
\usepackage{soul}
\usepackage{tabularx}
\usepackage{wrapfig}
\usepackage{enumitem}
\usepackage{adjustbox}
\usepackage{cleveref}
\usepackage{multirow}
\usepackage{array}
\usepackage{tikz}
\usetikzlibrary{arrows.meta,positioning,shapes.multipart,fit}
\usetikzlibrary{calc} 

\newcolumntype{L}[1]{>{\raggedright\arraybackslash}p{#1}}

\definecolor{exercisebgblue}{rgb}{0,  .69,  .941}
\definecolor{pastelviolet}{rgb}{.81,  .82,  .97}

\newcommand{\name}{\textsc{Sakura}\xspace}
\newcommand{\smalltt}[1]{\texttt{\small #1}}

\newcommand{\change}[1]{#1}                      

\newcounter{findingcounter}

\newenvironment{findingbox}[1]{%
\refstepcounter{findingcounter}%
\vspace{-3pt}%
\begin{tcolorbox}[
  colframe=pink!70!black, colback=pink!10, boxrule=.5pt, left=0pt, right=0pt, top=0pt, bottom=0pt 
]%
\textbf{Finding~\thefindingcounter:} #1%
}{%
\end{tcolorbox}%
\vspace{-7pt}%
}

\AtBeginDocument{%
  }

\setcopyright{cc}
\setcctype{by-nc-nd}
\acmDOI{10.1145/3832112}
\acmYear{2026}
\acmJournal{PACMSE}
\acmVolume{3}
\acmNumber{ISSTA}
\acmArticle{ISSTA021}
\acmMonth{10}
\acmSubmissionID{issta26main-p192-p}
\received{2026-01-30}
\received[accepted]{2026-04-16}

\begin{document}

\title{Sakura: An Approach for Generating Complex Tests from Natural Language Test Descriptions}

\author{Tyler Stennett}
\orcid{0009-0006-9780-9608}
\affiliation{%
  \institution{Georgia Institute of Technology}
  \city{Atlanta}
  \country{USA}
}
\email{tyler.stennett@gatech.edu}

\author{Rangeet Pan}
\orcid{0000-0002-8875-1225}
\affiliation{%
  \institution{IBM Research}
  \city{Yorktown Heights}
  \country{USA}
}
\email{Rangeet.Pan@ibm.com}

\author{Bridget McGinn}
\orcid{0009-0007-0442-9356}
\affiliation{%
  \institution{IBM Research}
  \city{Cambridge}
  \country{USA}
}
\email{Bridget.McGinn@ibm.com}

\author{Alessandro Orso}
\orcid{0000-0003-4516-9320}
\affiliation{%
  \institution{University of Georgia}
  \city{Athens}
  \country{USA}
}
\email{orso@uga.edu}

\author{Saurabh Sinha}
\orcid{0000-0003-4092-2643}
\affiliation{%
  \institution{IBM Research}
  \city{Yorktown Heights}
  \country{USA}
}
\email{sinhas@us.ibm.com}

\renewcommand{\shortauthors}{Stennett et al.}

\begin{abstract}
  Testing is a core activity in software development, and research on its automation has spanned several decades. Most existing approaches focus on generating unit tests for individual methods, validating isolated API endpoints, or targeting user interface (UI) layers. However, for non-API and non-UI tests, automated test generators typically exercise a single focal method. Recent empirical evidence~\cite{pan2025hamster} shows a substantial gap between such generated tests and developer-written tests, which often span several focal classes and methods, involve multi-step call sequences, and contain chained assertions—characteristics that current automated approaches fail to capture.
To address this gap, we propose generating tests from natural language (NL) descriptions of developer intent.
NL provides an expressive and accessible medium for specifying complex test scenarios and functional intent. We present \name, the first agent-based framework for generating structurally complex test cases from NL descriptions. \name decomposes NL descriptions into structured blocks and processes them using a multi-agent system consisting of a localization agent that grounds test steps in concrete application code via static analysis, a composition agent that synthesizes compilable test code and iteratively refines it using execution feedback, and a supervisor agent that coordinates agent interactions. 
To evaluate \name, we curate a novel dataset of NL test descriptions at three levels of abstraction, reflecting different end-user personas, systematically derived from developer-written tests in Apache Commons projects. Across 20 applications and 1,464 test scenarios, \name substantially outperforms off-the-shelf agentic tools such as Gemini CLI instantiated with multiple LLMs. Specifically, \name achieves 50--78\% higher test compilability and 38--66\% higher coverage overlap with ground-truth tests compared to baselines using the same models. Moreover, \name paired with small open-source models such as Devstral Small 2 and Qwen3-Coder outperforms Gemini CLI using large proprietary models, while also being more cost-effective.
\end{abstract}



\maketitle

\section{Introduction}
\label{sec:intro}

Software testing is critical for ensuring that applications behave as intended from the user’s perspective.
Achieving such validation in real-world systems, however, is often time-consuming and tedious, as it involves covering complex flows and interactions among different components. Developers must reason not only about individual functions or APIs, but also about operation sequences and application state evolution. Consequently, constructing a high-quality test suite for a realistic application demands significant manual effort.
To reduce this effort, decades of research~\cite{clarke1976testing, king1976symbolic, godefroid2005dart, sen2005cute, tillmann2008pex, harman:2010:tse, fraser2011evosuite, chen:2010:jss, lin:2009:ase, arcuri:2011:issta, lukasczyk:2023:emse, tufano2020unit, bareiss2022code, pizzorno2024coverup, pan2025aster} have explored automated test generation (ATG) across different testing levels, including unit~\cite{pasareanu:2010, godefroid2005dart, sen2005cute,  fraser2011evosuite, lin:2021, lukasczyk2022pynguin, pacheco2007feedback, ciupa:2008:icse, chen:2010:jss, lin:2009:ase, arcuri:2011:issta, lukasczyk:2023:emse}, integration~\cite{atlidakis_restler_2019, arcuri2019restful, viglianisi_resttestgen_2020, kim2023reinforcement, kim2023enhancing, kim_llamaresttest_2025, martin2022online, martin2021:blackandwhite, zhang_logiagent_2025}, and end-to-end~\cite{ui1, ui2, ui3, ui4, ui5}. 

\begin{figure}[t]
    \centering
    \includegraphics[width=\linewidth]{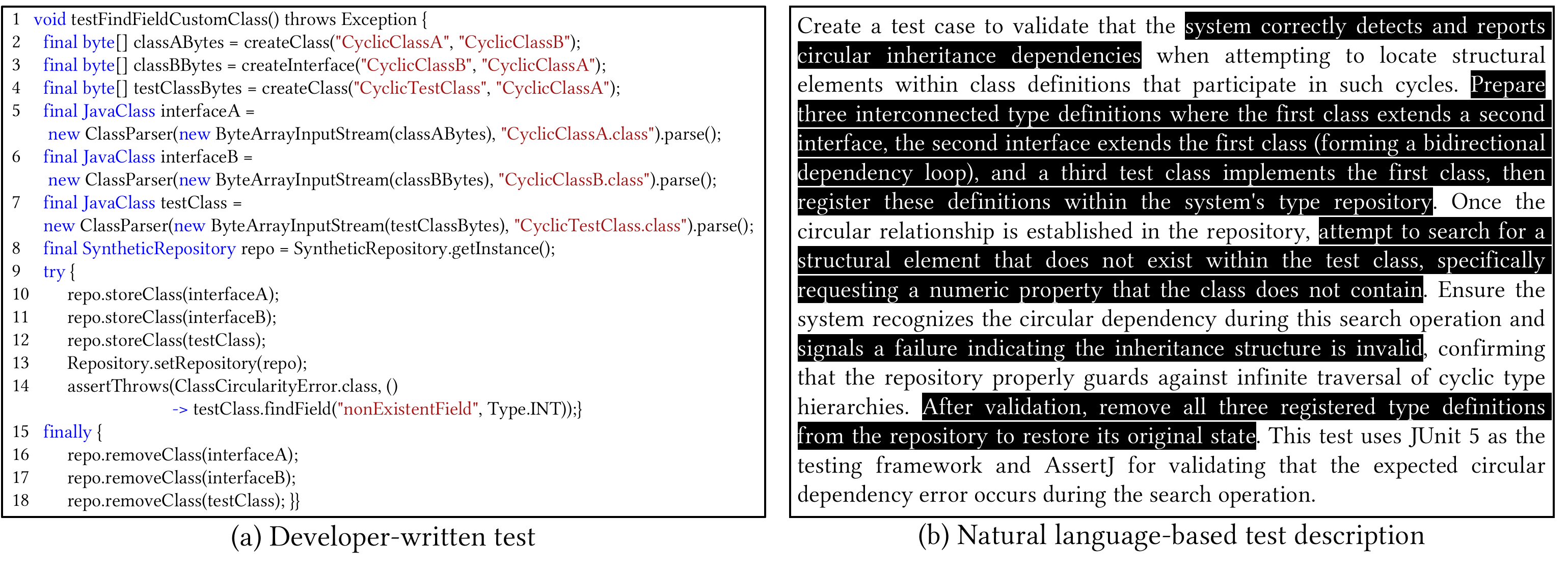}
    \caption{(a) A developer-written test from Apache Commons BCEL~\cite{apache_bcel} that checks circular inheritance dependencies; (b) the corresponding NL test description, with key logic elements highlighted.}
    \label{fig:motivation1}
\end{figure}

Despite this long history, existing ATG techniques have largely been designed around well-defined code components. Unit test generation typically targets individual methods or classes, API testing focuses on service endpoints, and end-to-end testing targets UI components. However, empirical studies~\cite{fraser:2015:developerstudy, pan2025aster, pan2025hamster} have shown that tests generated by automated approaches differ substantially from those written by developers along several dimensions, including naturalness~\cite{fraser:2015:developerstudy, pan2025aster} and structural complexity~\cite{pan2025hamster}. Although recent advances using large language models (LLMs) have helped address issues of test naturalness~\cite{pan2025aster} by leveraging knowledge learned from developer-written tests, the challenge of capturing structural complexity remains largely unaddressed.

A recent large-scale empirical study~\cite{pan2025hamster} characterizes this gap. Focusing on non-API and non-UI tests, the study observed that developer-written tests frequently span multiple classes and methods, beyond the conventional definition of unit tests centered on a single focal method.
\change{This breadth of behavior is reflected in the tests themselves, which often exhibit non-trivial setup logic, extensive and interdependent call sequences, lengthy assertion chains, and explicit teardown behavior~\cite{pan2025hamster}. We refer to such tests as \emph{structurally complex}. Our evaluation quantifies test complexity by focal-method count, which captures this behavioral breadth directly, independent of how any particular test implements it (\S\ref{subsubsec:rq5}).}
\change{In contrast to these complex tests, state-of-the-art} ATG tools such as EvoSuite~\cite{fraser2011evosuite} and \textsc{ASTER}~\cite{pan2025aster} are primarily designed to operate within a single-class, single-method scope, limiting their ability to reflect the structure and intent of developer-written tests.

Figure~\ref{fig:motivation1} illustrates both this complexity and a potential path forward using an example from the Apache BCEL project~\cite{apache_bcel}. Figure~\ref{fig:motivation1}(a) shows a developer-written test that exercises multiple classes and methods through a sequence of invocations with explicit teardown logic. To the best of our knowledge, no existing ATG tool can generate tests with comparable complexity or capture the underlying developer intent. Figure~\ref{fig:motivation1}(b) shows a natural language (NL) description of the test, specifying the intended behavior: constructing circular inheritance dependencies, performing a lookup for a non-existent element, asserting the expected failure, and defining teardown behavior, along with details about the testing framework and assertion library.

\begin{figure}[t]
    \centering
    \includegraphics[width=\linewidth]{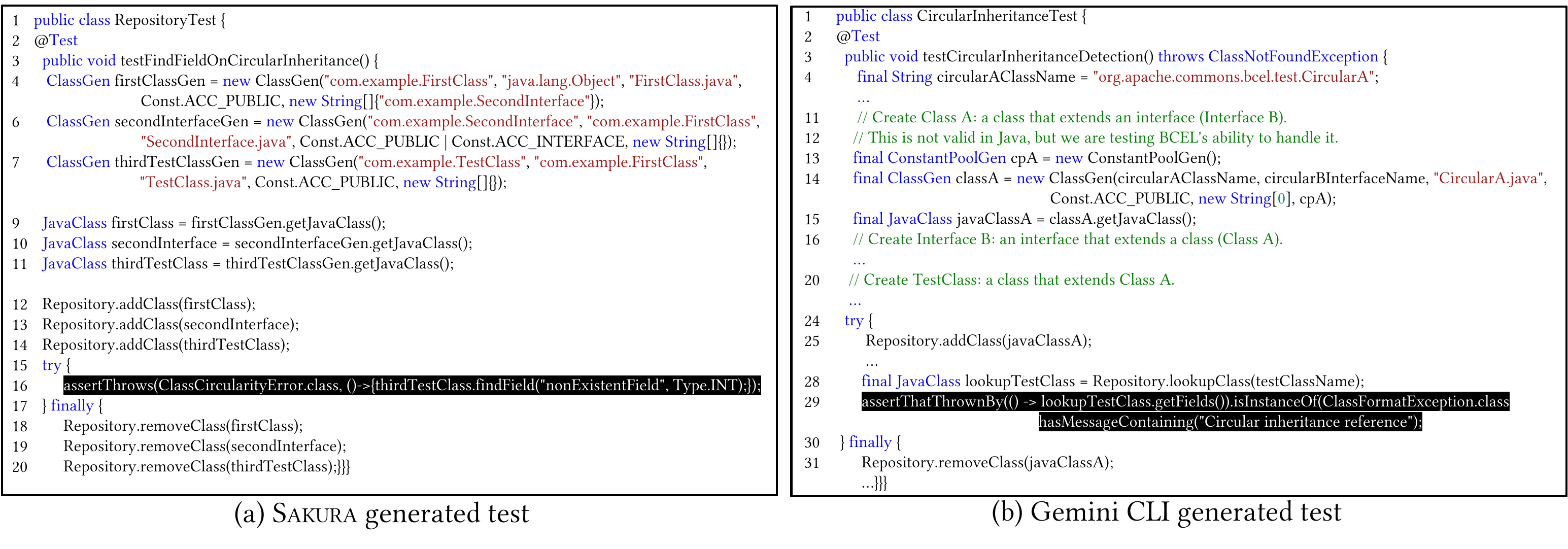}
     \caption{The test generated by (a) \name and (b) Gemini CLI for the NL description in Figure~\ref{fig:motivation1}(b). The highlighted region illustrates the key differences in the generated tests.}
    \label{fig:motivation2}
\end{figure}

This pairing suggests an alternative to code-centric test generation: starting from intent expressed as NL-based test descriptions. \change{Such descriptions arise naturally in practice---for example, as the intent expressed in a developer request for a test, a scenario to be validated, or reproduction steps in an issue report---and provide an expressive way to specify complex behaviors, spanning multiple components and interactions, that code-centric techniques cannot target.} Recent advances in LLMs, including agentic approaches~\cite{gemini_cli, claude_code, ibm:wca4eja}, have shown promise in generating tests from NL descriptions. However, due to their broad, unrefined toolsets, these approaches typically lack the structural awareness of the application necessary to realize the intended behavior and produce compilable tests. Figure~\ref{fig:motivation2} illustrates this limitation by comparing tests generated from the description in Figure~\ref{fig:motivation1}(b) by Gemini CLI~\cite{gemini_cli} and by our approach, \name, both instantiated with the same LLM (Gemini~2.5~Pro). Although Gemini CLI correctly follows several high-level steps---such as creating the circular dependency and using the appropriate testing framework---it fails to localize the correct application methods and exceptions. Specifically, it invokes \smalltt{getFields} instead of \smalltt{findField}, which does not perform inheritance-aware lookup and cannot trigger the intended error, and asserts \smalltt{ClassFormatException} rather than the expected \smalltt{ClassCircularityError}. In contrast, \name generates a test with the correct call sequence and assertions, faithfully reflecting the behavior specified in the description.

This difference arises from the underlying architecture. \name adopts a novel multi-agent design that takes an NL test description as input and generates compilable and executable tests as output. The approach is explicitly designed to support structurally complex tests that span multiple classes and methods. \name begins by decomposing the NL description into a structured intermediate representation. Inspired by behavior-driven development (BDD)~\cite{bdd}, the input is transformed into a set of BDD-style blocks that make the intended test behavior explicit. Based on this representation, \name employs three specialized agents: (1) a \emph{localization agent} that uses static analysis to identify relevant classes and methods, (2) a \emph{composition agent} that integrates static analysis and code execution to synthesize executable tests, and (3) a \emph{supervisor agent} that coordinates agent interactions and manages auxiliary tasks. The generated tests are written to the designated test directory. In the current implementation, \name targets Java applications.

Beyond the proposed approach, we contribute a curated evaluation dataset for studying the generation of structurally complex tests. We select widely used projects from the Apache Commons repository~\cite{apache} and identify developer-written tests, retaining only those introduced after the cutoff dates of the models used in our evaluation to mitigate data leakage.
For each selected test, we generate NL descriptions at multiple abstraction levels, ranging from implementation-level detail to high-level intent. \change{Because no existing corpus pairs practitioner-authored test descriptions with ground-truth tests at scale, these generated descriptions serve as approximations of the real user inputs described above; they are produced by a state-of-the-art LLM distinct from the models under evaluation.} The resulting dataset consists of 1,464 NL descriptions paired with 488 developer-written tests drawn from 20 Apache projects. We publicly release the dataset~\cite{artifact}.

We evaluate \name against two closed-source models, Gemini 2.5 Pro and Gemini 2.5 Flash~\cite{comanici2025gemini}, and two open-source models, Qwen3-Coder~\cite{qwen3} and Devstral Small 2~\cite{devstral}. The generated tests are compared against developer-written tests, which serve as ground truth, and against tests produced by off-the-shelf LLM-based agentic tools, such as Gemini CLI, with two compatible LLMs. Our results demonstrate that \name substantially outperforms off-the-shelf agentic baselines across all evaluated dimensions, achieving 50--78\% higher compilability, 38--66\% greater coverage overlap, and 31--72\% improved structural fidelity when using the same models. These gains are more notable on challenging inputs, with \name outperforming baselines by up to 140\% on complex tests with multiple focal methods. Notably, \name with open-source models consistently surpasses Gemini CLI with proprietary models at equal or lower cost, demonstrating that our structured multi-agent workflows can compensate for model capacity. Our key contributions include:
\begin{itemize}[leftmargin=*]
    \item A novel multi-agent approach for generating structurally complex Java tests from natural language test descriptions that existing ATG techniques cannot target.
    \item Comprehensive empirical evaluation of \name against state-of-the-art LLM-based agentic coding approaches across multiple large language models, including both closed-source and open-source models, showing that smaller open-source models can outperform off-the-shelf agentic approaches that rely on larger models.
    \item A curated dataset consisting of 1,464 (test description, test) pairs, along with a fully automated evaluation pipeline to support reproducible assessment of intent-driven and complex test generation approaches. The dataset and artifacts are publicly available~\cite{artifact}.
\end{itemize}

\vspace{-2pt}
\section{\name}
\label{sec:approach}

\begin{figure}[t]
    \centering
    \includegraphics[width=\linewidth]{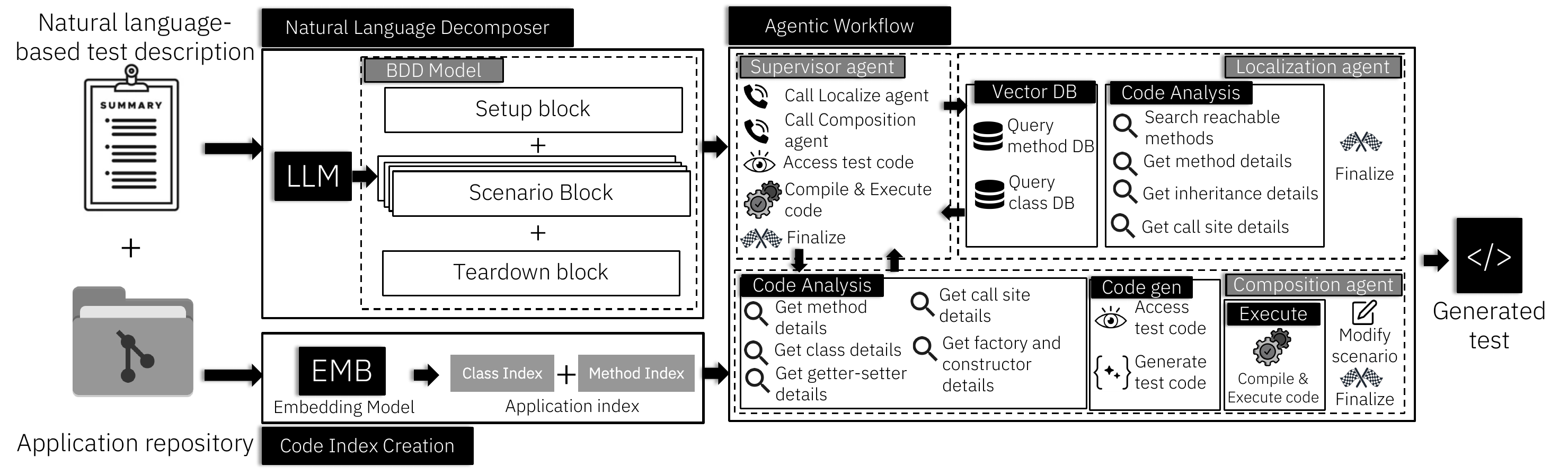}
    \caption{Overview of \name (EMB: Embedding Model, BDD: Behavior-Driven Development).} 
    \vspace{-5pt}
    \label{fig:nl2test}
\end{figure}

Figure~\ref{fig:nl2test} presents an overview of \name. Given a test description and an application, \name produces a compilable and executable test through three phases: (1)~\emph{code index creation}, a one-time preprocessing step that embeds the codebase into a searchable vector space; (2)~the \emph{NL decomposer}, which converts the test description into a structured BDD-style representation; and (3)~an \emph{agentic workflow} that combines localization, composition, and supervisor agents to synthesize the test. \change{Throughout this section, we use the circular inheritance test from Figure~\ref{fig:motivation1} as a running example.}

\vspace{-3pt}
\subsection{Code Index Creation}
\label{subsec:codeindex}

In the first step, we construct searchable vector indices over the target project’s codebase. These indices support the subsequent agentic workflow by enabling the retrieval of relevant code elements and facilitating LLMs’ understanding of the application’s structure and semantics.
We index two types of code elements—methods and classes—to support flexible retrieval. For method indexing, we traverse each class and extract all application methods, including inherited methods that are visible through the class hierarchy. Each method is indexed together with its enclosing class declaration to preserve contextual information necessary for disambiguation. We explicitly track both the declaring class, in which the method is originally defined, and the containing class, through which the method is accessed, allowing inheritance relationships to be correctly resolved (statically) during retrieval. For class indexing, we extract all non-test classes that expose at least one publicly visible method. In both cases, we apply filtering to exclude test artifacts.
The code indices are constructed once per project and reused across all test descriptions for that project. 

\vspace{-12pt}
\subsection{Natural Language Decomposer}
\label{subsec:decomposeNL}

With the code indices in place, we next process individual test descriptions. This stage transforms unstructured NL into a structured intermediate representation suitable for code localization and test synthesis. Instead of directly mapping free-form text to executable code, which can overwhelm the LLM and lead to poor localization, we introduce an intermediate representation that preserves semantic intent while imposing structural constraints to guide subsequent processing.

\begin{wrapfigure}{r}{0.54\columnwidth}
    \centering
    \vspace{-6pt}
    \includegraphics[width=\linewidth]{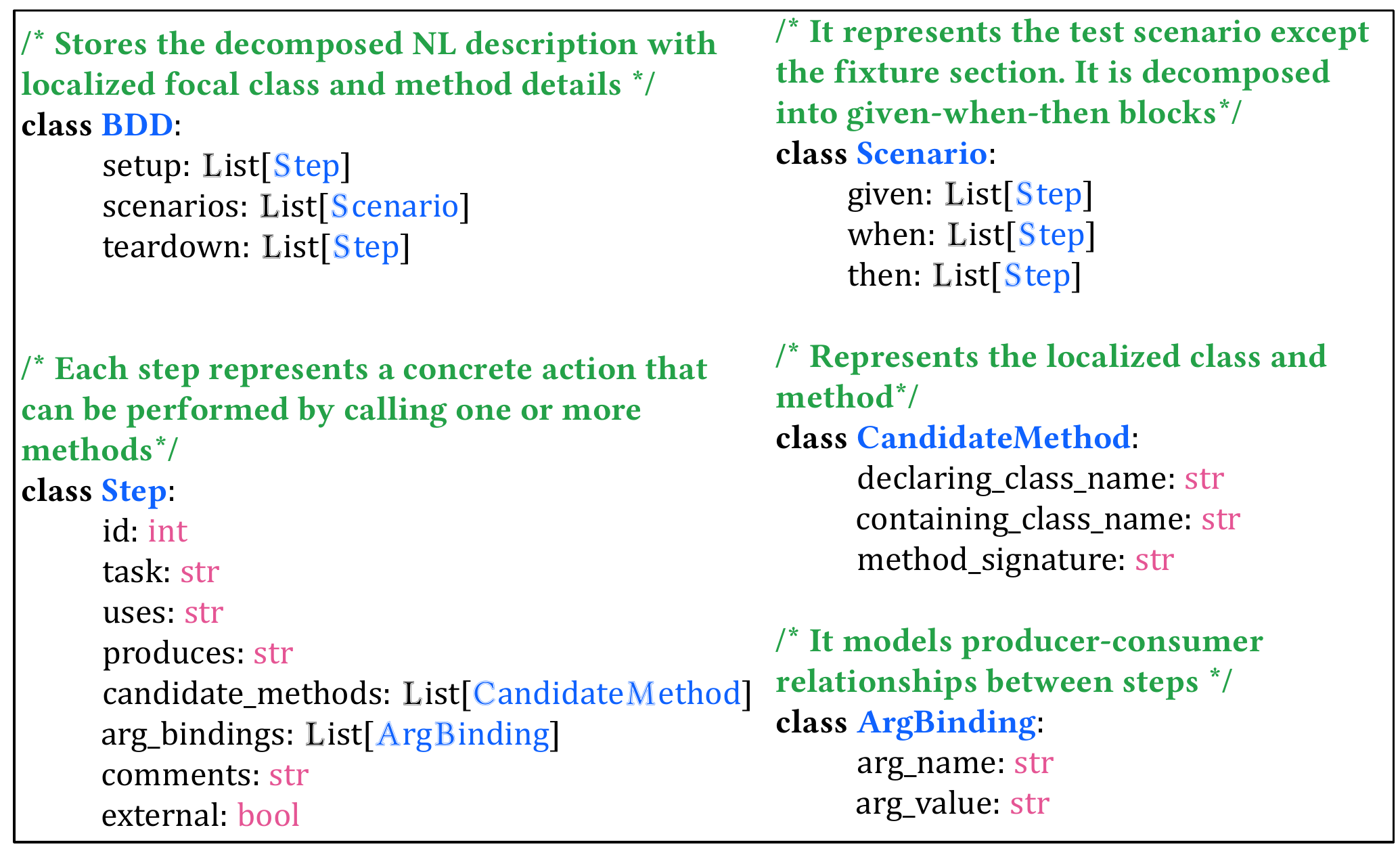}
    \caption{The BDD model.}
    \label{fig:decomp-model}
    \vspace{-7pt}
\end{wrapfigure}

We adopt a BDD-inspired~\cite{bdd} decomposition model, depicted in Figure~\ref{fig:decomp-model}, that organizes test behavior into setup steps, one or more scenarios, and teardown steps; each scenario is further divided into \emph{given}, \emph{when}, and \emph{then} blocks, encoding preconditions, actions, and postconditions, respectively. The model accommodates fixtures, cleanly separates preconditions from actions and assertions, and, by permitting multiple scenarios, supports complex tests that span multiple application features whose coverage and validation may be interwoven in the original description.
Each \smalltt{Step} captures a discrete testing action through an NL task description, along with \smalltt{uses} and \smalltt{produces} fields that explicitly track data flow between steps, auxiliary NL annotations, and an \smalltt{external} flag indicating whether the step involves application code or externally imported libraries.

The NL decomposer constructs this initial decomposition by prompting an LLM to conform to the schema and divide the description into its most atomic form. \change{Its output is programmatically validated against the schema, and any violations (e.g., missing fields or malformed step references) are returned to the model as feedback for a corrected decomposition, following the error-guided repair strategy established in prior work~\cite{gou2024critic}. We allow up to three repair attempts; in our evaluation (\S\ref{sec:evaluation}), no run exhausted this budget for any model or test description. The validated decomposition is then passed to the agentic workflow, where the localization agent grounds each step to concrete application code, augmenting it with \smalltt{CandidateMethod} blocks that record the localized classes and methods and \smalltt{ArgBinding} blocks that capture inter-step data dependencies.}

\begin{figure}[t]
    \centering
    \includegraphics[width=0.86\linewidth]{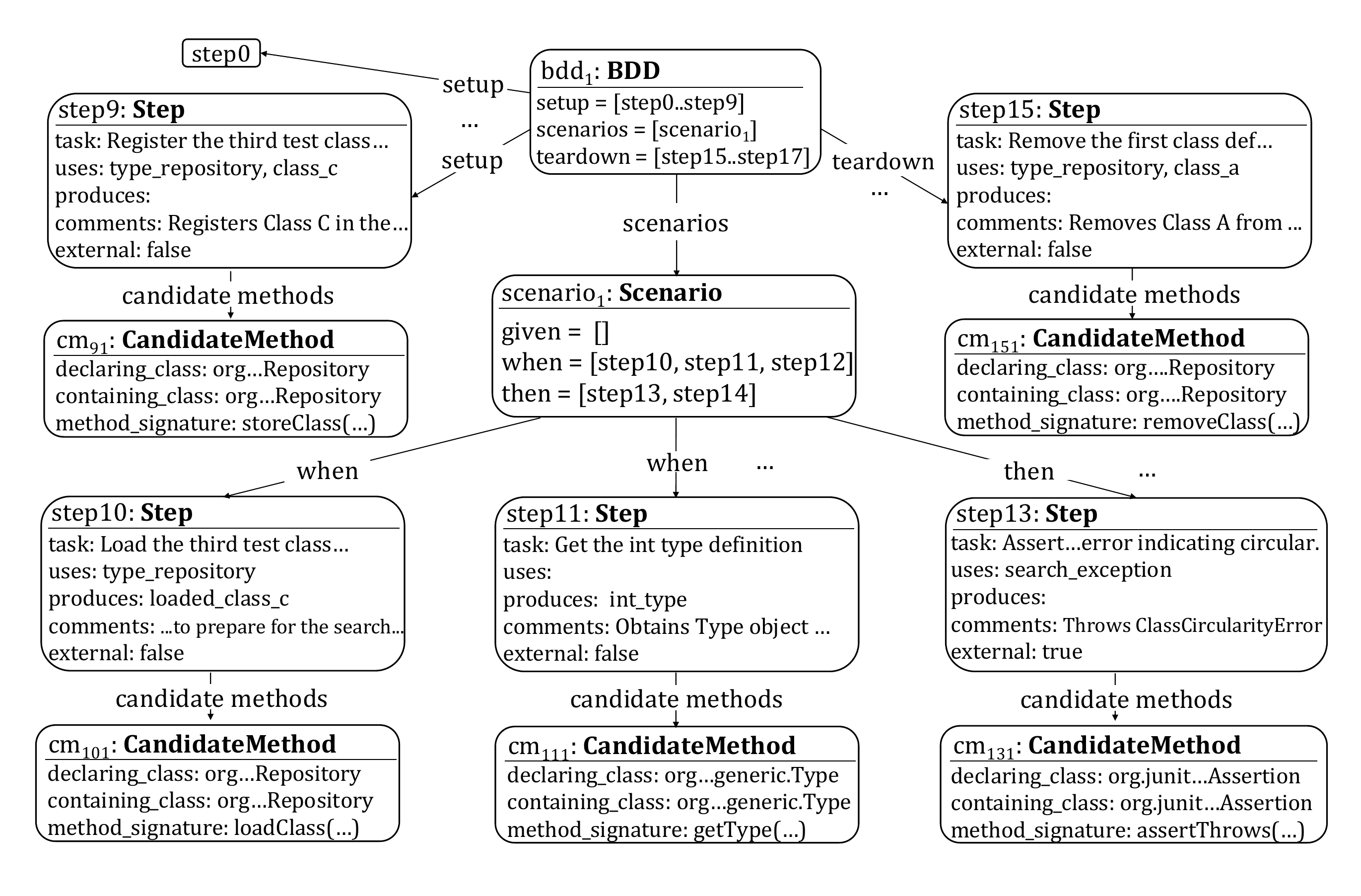}
    \caption{The pruned decomposition model for the circular inheritance detection test scenario (Figure~\ref{fig:motivation1}). \change{Refer to our artifact~\cite{artifact_motivating} for the complete decomposition model.}} 
    \label{fig:decomposition-example}
\end{figure}

Figure~\ref{fig:decomposition-example} illustrates the localized decomposition for the \change{running example} shown in Figure~\ref{fig:motivation1}. The complex test scenario is broken down into a sequence of fine-grained steps: for instance, \smalltt{Step13} asserts that a circularity error is raised and has its \smalltt{external} flag set, as it interacts with the testing library rather than application classes. The \smalltt{CandidateMethod} blocks shown beneath each step reflect the results of this localization; for clarity, we omit \smalltt{ArgBinding} blocks from the figure.

\vspace{-8pt}
\subsection{Agentic Workflow}
\label{subsec:agents}

Having constructed the code indices and the BDD representation, we employ a hierarchical multi-agent system to generate tests. This design reduces prompt complexity by assigning each subagent a focused responsibility and toolset, while enabling adaptive control flow where the supervisor can reroute execution based on intermediate results. The system comprises three agents: a \emph{localization agent} that grounds each BDD step in concrete application code, a \emph{composition agent} that assembles localized steps into a complete test through iterative generation and validation, and a \emph{supervisor agent} that coordinates execution and routes failures to the appropriate subagent for correction.

\textit{Communication Protocol.}
Each agent is initialized with a curated system prompt and task-specific inputs, and invokes tools until completion criteria are met. To improve token efficiency, subagents do not maintain persistent state; instead, the structured BDD representation and evolving test code are passed between agents, with the supervisor providing summaries and targeted instructions.

\vspace{-4pt}
\subsubsection{Localization Agent}
\label{subsubsec:localize}

The localization agent maps each block in the structured output produced by the NL decomposer to relevant classes and methods. Starting from the decomposed blocks and the original NL test description, the agent iteratively queries the code indices and applies static analysis to identify and validate candidate implementations.
For each block, the agent augments the decomposition model (Section~\ref{subsec:decomposeNL}) with a set of candidate methods, specified by fully qualified class names and method signatures, and records annotations capturing localization rationale or uncertainty. When a block involves external dependencies (e.g., library code), the agent marks it with an external flag and infers the relevant classes and methods using domain knowledge.

\change{For the running example, the localization agent grounds the step ``requesting a numeric property that the class does not contain'' to \smalltt{JavaClass.findField}. By inspecting the method's implementation, the agent determines that it delegates to a hierarchy-aware helper that raises \smalltt{ClassCircularityError}---the failure that ``signals \ldots the inheritance structure is invalid,'' which the description's assertion step expects---and thereby distinguishes \smalltt{findField} from similarly named accessors, such as \smalltt{getFields}, that do not traverse the class hierarchy (Figure~\ref{fig:motivation2}). The complete localized decomposition for this example is available in our artifact~\cite{artifact_motivating}.}

\begin{table}[!htp]
\centering
\footnotesize
\setlength{\tabcolsep}{4pt}
\renewcommand{\arraystretch}{1.15}
\caption{Localization agent tools.}
\begin{tabular}{>{\centering\arraybackslash}m{0.055\linewidth}@{}l@{}}
\toprule
& \begin{tabular}[c]{@{}L{0.27\linewidth}L{0.62\linewidth}@{}}
  \textbf{Tool} & \textbf{Purpose}
  \end{tabular} \\
\midrule
\rotatebox{90}{Semantic Tools} &
\begin{tabular}[c]{@{}L{0.27\linewidth}L{0.62\linewidth}@{}}
\texttt{method\_vector\_search} &
Returns the top-$k$ methods semantically similar to an NL query, along with their qualified declaring and implementing class names. \\
\texttt{class\_vector\_search} &
Returns the top-$k$ classes most semantically similar to a query, supporting broader exploration when the relevant class is unknown or method-level search is insufficient. \\
\texttt{reachable\_methods\_search} &
Constrains semantic search to methods callable from a specified class, retrieved through static analysis, with configurable visibility filtering.
\end{tabular} \\
\midrule
\rotatebox{90}{\parbox{2.2cm}{\centering Static Analysis Tools}} &
\begin{tabular}[c]{@{}L{0.27\linewidth}L{0.62\linewidth}@{}}
\texttt{get\_method\_details} &
Returns structured metadata for a method, including parameter types, return type, modifiers, and annotations. \\
\texttt{get\_inherited\_\allowbreak external\_types} &
Lists external types in a class's inheritance chain, supporting reasoning about superclasses and potentially inherited methods not accessible through index search. \\
\texttt{get\_call\_site\_details} &
Extracts metadata for all callees within a method body, including names, return types, parameters, and modifiers.
\end{tabular} \\
\midrule
\rotatebox{90}{\parbox{0.45cm}{\centering Code Tool}} &
\begin{tabular}[c]{@{}L{0.27\linewidth}L{0.62\linewidth}@{}}
\texttt{extract\_method\_source} &
Retrieves source code for a candidate method with optional line range specification.
\end{tabular} \\
\bottomrule
\end{tabular}
\label{tab:localizationtools}
\end{table}

\textit{Tool Suite.}
Table~\ref{tab:localizationtools} summarizes the localization tools, organized into three categories: semantic retrieval tools for querying vector indices, static analysis tools for reasoning about signatures, inheritance, and visibility, and code inspection tools for selectively accessing source code when the metadata is insufficient.

\vspace{-4pt}
\subsubsection{Composition Agent}
\label{subsubsec:composition}

The role of the composition agent is to synthesize compilable and executable Java test code using the localization results and additional feedback, such as compilation and execution outcomes. Unlike the localization agent, which focuses on grounding test steps to relevant code elements, the composition agent is primarily responsible for code synthesis and iterative validation, and operates through a sequence of specialized tools.
Given the localized decomposition, the original test description, and any feedback provided by the supervisor, the composition agent follows a feedback-driven refinement loop. It first inspects the localized candidates to assess type compatibility and determine appropriate object instantiation patterns. It then generates an initial test implementation and validates it through compilation and execution. When errors arise---such as type mismatches, missing imports, or assertion failures---the agent examines the affected code regions, identifies the underlying root causes, and applies targeted corrections. This closed-loop process anchors test generation in concrete build-system feedback, rather than relying solely on LLM-inferred assumptions about Java semantics or project structure.

\change{For the running example, this loop uses compilation feedback to resolve issues such as incorrect package placement and missing imports (e.g., for the \smalltt{Const} type). The Gemini CLI baseline lacks such feedback: beyond the localization errors highlighted in Figure~\ref{fig:motivation2}(b), its generated test fails to compile due to similar import and placement defects (omitted from the figure; see our artifact~\cite{artifact_motivating}).}

\begin{table}[!htp]
\centering
\footnotesize
\setlength{\tabcolsep}{4pt}
\renewcommand{\arraystretch}{1.15}
\caption{Composition agent tools (Decomp: Decomposition, Gen: Generation).}
\begin{tabular}{>{\centering\arraybackslash}m{0.055\linewidth}@{}l@{}}
\toprule
& \begin{tabular}[c]{@{}L{0.27\linewidth}L{0.62\linewidth}@{}}
  \textbf{Tool} & \textbf{Purpose}
  \end{tabular} \\
\midrule
\rotatebox{90}{Inspection Tools} &
\begin{tabular}[c]{@{}L{0.27\linewidth}L{0.62\linewidth}@{}}
\texttt{get\_method\_details} &
Described in Table~\ref{tab:localizationtools}. \\
\texttt{get\_call\_site\_details} &
Described in Table~\ref{tab:localizationtools}. \\
\texttt{extract\_method\_source} &
Described in Table~\ref{tab:localizationtools}. \\
\texttt{get\_class\_fields} &
Lists all field declarations for a class, including names, types, and modifiers.
\end{tabular} \\
\midrule
\rotatebox{90}{State Tools} &
\begin{tabular}[c]{@{}L{0.27\linewidth}L{0.62\linewidth}@{}}
\texttt{get\_getters\_and\_setters} &
Identifies property accessor methods for a class, enabling the agent to manipulate object state during test setup and extract values for assertions. \\
\texttt{get\_constructors\_\allowbreak and\_factories} &
Enumerates constructors and static factory methods, allowing the agent to identify appropriate instantiation patterns.
\end{tabular} \\
\midrule
\rotatebox{90}{Project Tool} &
\begin{tabular}[c]{@{}L{0.27\linewidth}L{0.62\linewidth}@{}}
\texttt{extract\_project\_\allowbreak dependencies} &
Retrieves dependencies from the build configuration (e.g., Maven POM or Gradle build files), informing the agent about available testing frameworks, assertion libraries, and mocking utilities. This ensures generated tests align with the project's existing test infrastructure.
\end{tabular} \\
\midrule
\rotatebox{90}{Gen. and Validation Tools} &
\begin{tabular}[c]{@{}L{0.27\linewidth}L{0.62\linewidth}@{}}
\texttt{generate\_test\_code} &
The primary synthesis tool, accepting raw Java source code along with the target class name and test method signature. The system parses the submission, determines the appropriate location within the test directory structure based on the declaring package, and writes the file to disk. \\
\texttt{compile\_and\_execute\_test} &
Invokes the project's build system to compile and run the generated test. Returns structured feedback including compilation errors with line numbers and messages, runtime exceptions, and assertion failure details. \\
\texttt{view\_test\_code} &
Retrieves the current contents of the generated test file, optionally restricted to specific line ranges.
\end{tabular} \\
\midrule
\rotatebox{90}{Decomp. Tool} &
\begin{tabular}[c]{@{}L{0.27\linewidth}L{0.62\linewidth}@{}}
\texttt{modify\_step\_comment} &
Updates the comment field of a specific step in the decomposition model, allowing the agent to record insights discovered during composition (e.g., incorrect candidate methods or missing type information).
\end{tabular} \\
\bottomrule
\end{tabular}
\label{tab:compositiontools}
\end{table}

\textit{Tool Suite.}
Table~\ref{tab:compositiontools} summarizes the composition tools, organized into five categories: metadata and code inspection, state manipulation, project context, code generation and validation, and decomposition feedback. The code generation and validation tools form the core refinement loop: the agent generates test code, invokes the build system, and retrieves structured feedback---compilation errors, runtime exceptions, and assertion failures---to guide corrections. Decomposition feedback tools allow the agent to annotate the BDD model with code composition insights, enabling the supervisor to trigger re-localization if needed and informing the localization agent of any concerns to address.

\subsubsection{Supervisor Agent}
\label{subsubsec:supervisor}
The supervisor agent coordinates the workflow by managing task delegation, information flow, and termination decisions. Both subagents return structured outputs: the localization agent provides the augmented decomposition with candidate methods and annotations summarizing ambiguities or unresolved dependencies; the composition agent returns the generated test code along with execution status and feedback on localization adequacy.

Using these outputs, the supervisor maintains the evolving decomposition state and decides whether to advance, retry, or terminate. Central to this role is failure diagnosis: the supervisor distinguishes composition-level issues (e.g., incorrect instantiation patterns) from localization-level issues (e.g., incompatible method signatures) and re-invokes the appropriate agent with focused instructions. Before termination, the supervisor independently verifies the generated test by compiling and executing it, ensuring alignment with the test description.

\textit{Tool Suite.}
Table~\ref{tab:supervisortools} summarizes the supervisor's toolset: orchestration tools for delegating structured inputs to subagents, and verification tools for compiling and executing generated tests. Restricting the supervisor to these two categories keeps it focused on coordination, leaving all code manipulation to the subagents.

\begin{table}[t]
\centering
\footnotesize
\setlength{\tabcolsep}{4pt}
\renewcommand{\arraystretch}{1.15}
\caption{Supervisor agent tools (Verifi: Verification).}
\begin{tabular}{>{\centering\arraybackslash}m{0.055\linewidth}@{}l@{}}
\toprule
& \begin{tabular}[c]{@{}L{0.27\linewidth}L{0.62\linewidth}@{}}
  \textbf{Tool} & \textbf{Purpose}
  \end{tabular} \\
\midrule
\rotatebox{90}{\parbox{2.2cm}{\centering Orchestration Tools}} &
\begin{tabular}[c]{@{}L{0.27\linewidth}L{0.62\linewidth}@{}}
\texttt{call\_localization\_agent} &
Delegates to the localization agent with targeted instructions and the current decomposition state. When refinements apply to specific steps, they are explicitly identified to avoid unnecessary tool invocations and reduce token usage. \\
\texttt{call\_composition\_agent} &
Delegates to the composition agent with implementation guidance, the localized decomposition, and curated feedback from previous composition attempts. This feedback enables the agent to avoid repeating past failures.
\end{tabular} \\
\midrule
\rotatebox{90}{Verifi. Tools} &
\begin{tabular}[c]{@{}L{0.27\linewidth}L{0.62\linewidth}@{}}
\texttt{compile\_and\_execute\_test} &
Described in Table~\ref{tab:compositiontools}.\\
\texttt{view\_test\_code} &
Described in Table~\ref{tab:compositiontools}.
\end{tabular} \\
\bottomrule
\end{tabular}
\label{tab:supervisortools}
\end{table}

\section{\name Evaluation Dataset}
\label{sec:dataset}

This section describes the dataset constructed to assess \name and other agentic test generation approaches. \change{In practice, NL inputs for test generation can come from artifacts like developer-authored descriptions of intended test behavior, requirement-style scenarios to be validated, and issue reports. To our knowledge, however, no corpus pairs such inputs with ground-truth tests at the scale our evaluation requires.} We therefore select high-quality Java applications, mine developer-written tests, and generate NL descriptions that approximate these user inputs, capturing test intent at varying levels of abstraction. The dataset curation process consists of three stages. 

\begin{figure}[t]
    \centering
    \includegraphics[width=0.91\linewidth]{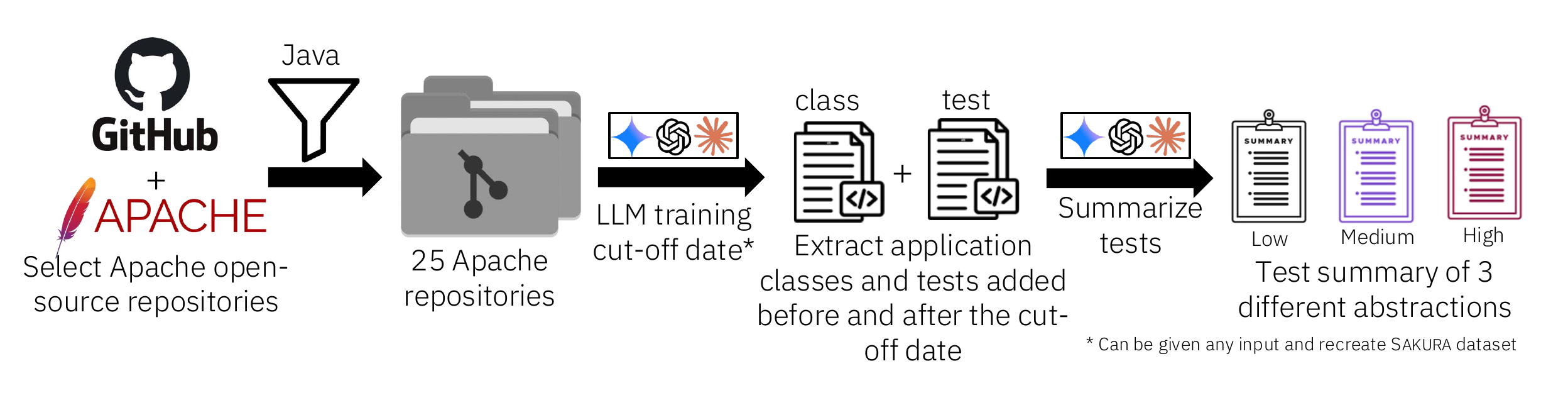}
    \caption{Overview of \name evaluation dataset creation steps.} 
    \label{fig:test2nl}
\end{figure}

First, in the \emph{repository selection and filtering} stage, we identify mature, widely used Java projects and select developer-written tests that are introduced after the training cutoff dates of the evaluated LLMs, thereby minimizing data leakage. Second, during \emph{context extraction}, we perform static analysis on the selected applications to extract the information required to accurately describe test behavior, including relevant classes, methods, and interactions. Finally, in the \emph{abstraction-based description generation} stage, we use Claude Sonnet 4.5~\cite{anthropic2025claude45}---a different model from those evaluated in our later experiments (\S\ref{subsec:setup})---to generate descriptions at three levels of abstraction. These range from low-level descriptions that include substantial implementation detail, suitable for developers actively working on the codebase, to high-level, scenario-oriented descriptions that abstract away implementation details and are more appropriate for product owners. To match the nature of each abstraction level, we vary the generation temperature: 0.3 for low-level descriptions requiring precision, 0.5 for medium-level, and 0.7 for high-level descriptions, where more natural, varied phrasing is desirable~\cite{renze2024effect}. Figure~\ref{fig:test2nl} depicts these steps. 

\subsection{Repository Selection and Filtering}

Our curation pipeline begins by selecting repositories that offer comprehensive test suites, active maintenance, and widespread adoption. We initially selected 25 Apache Commons open-source projects and verified that both application code and test suites compile successfully. 
From these projects, we extract test methods through a multi-stage filtering process using JavaParser~\cite{javaparser} for static analysis and the Hamster~\cite{pan2025hamster} test identification methodology, which uses testing frameworks, annotations, and naming conventions to identify test methods.
We then apply a series of filters. First, we use Hamster’s focal method identification to retain only tests that exercise application behavior, excluding tests without application focal methods (e.g., library-only tests). Second, we remove non-test constructs, including abstract classes, interfaces, enums, and annotation declarations. Third, we exclude tests annotated with \smalltt{@Disabled}, which may target deprecated behavior, and \smalltt{@TestFactory}, which generate tests dynamically and fall outside the scope of our evaluation.
Finally, to mitigate data contamination from tests that may appear in LLM training corpora, we apply a temporal filtering step. A test is retained only if it was added after the latest training cutoff among the evaluated models (\S\ref{subsec:setup}).

\subsection{Context Extraction}
\label{subsec:context_ext}

In this step, we use static analysis to extract the code context needed to accurately characterize test behavior. For each test method, we collect contextual information across five categories: (1)~\emph{test class context}, including field declarations and class-level annotations; (2)~\emph{fixture methods}, capturing setup and teardown logic with their invocations, variable declarations, source code, and Javadoc comments; (3)~\emph{test method} details, including invocations, declarations, source code, and annotations encoding execution semantics (e.g., expected exceptions and parameterization); (4)~\emph{helper methods}, identified as non-test methods in the test suite, invoked by fixture or test methods, expanded to depth one and inlined to ensure self-contained descriptions; and (5)~\emph{application context}, covering all exercised classes and methods with their signatures, inheritance relationships, and documentation.

\subsection{Abstraction-Based Description Generation}
\label{subsec:abstraction_desc}

\begin{table}[t]
\centering
\footnotesize
\caption{Abstraction level criteria for the \name evaluation dataset.}
\begin{tabularx}{\linewidth}{>{\raggedright\arraybackslash}p{1.9cm} >{\raggedright\arraybackslash}X >{\raggedright\arraybackslash}X >{\raggedright\arraybackslash}X}
\toprule
\textbf{Criterion} &
\textbf{Low Abstraction} &
\textbf{Medium Abstraction} &
\textbf{High Abstraction} \\
\midrule
Identifier Specificity &
Uses exact class, method, and variable names from the codebase. &
Uses semantic descriptors for classes and methods; variable names only when required for disambiguation. &
Prohibits technical identifiers; uses business-level entities and concepts only. \\
\addlinespace[2pt]
Helper Treatment &
Helpers are fully unwrapped; all helper logic is inlined step by step. &
Helpers are described architecturally by intent without implementation details. &
Helpers are invisible; only resulting states or preconditions are described. \\
\addlinespace[2pt]
Application Code Internals &
Specifies exact method invocations and arguments, omitting internal behavior. &
Treats application code as a black box, describing invocation and observable effects. &
Focuses solely on observable business outcomes without method-level references. \\
\addlinespace[2pt]
Method Chaining &
Enumerates each chained call explicitly and sequentially. &
Collapses chains into intent-based logical operations. &
Ignores chaining entirely; describes only the final state. \\
\addlinespace[2pt]
Inputs and Data &
Requires exact literal values for all inputs. &
Describes inputs by type, constraints, or characteristics. &
Uses high-level domain archetypes and scenarios. \\
\addlinespace[2pt]
Assertions &
Lists every assertion with exact APIs and expected values. &
Describes assertion intent and references relevant variables. &
Frames verification in business terms without technical details. \\
\bottomrule
\end{tabularx}
\label{tab:abstractions}
\end{table}
The final dataset component comprises test descriptions at three abstraction levels---\emph{low}, \emph{medium}, and \emph{high}---corresponding to progressively reduced implementation detail. \change{This design enables evaluation under varying input specificity, mirroring the range of inputs users would author in practice: low-level descriptions approximate implementation-oriented test intent written by developers actively working on the codebase, whereas high-level descriptions approximate scenario-oriented inputs such as story-like requirements.}

Each level is realized through a distinct system prompt assigning the LLM a persona (\emph{Senior Technical Lead}, \emph{Software Architect}, or \emph{Technical Product Owner}) and defining how content categories should be expressed. Six criteria, summarized in Table~\ref{tab:abstractions}, govern how each category is included, summarized, or omitted. These criteria collectively define a systematic gradient from implementation-complete specifications (low), through architectural guidance permitting flexibility (medium), to business-level requirements demanding domain knowledge and technical inference (high). For instance, \emph{identifier specificity} ranges from exact class and method names (low) to semantic descriptors like ``the payment service'' (medium) to pure business-level concepts (high). Similarly, \emph{helper treatment} progresses from fully inlined step-by-step logic, through architectural descriptions of helper intent, to complete omission where only the resulting state is visible. \change{To illustrate this gradient concretely, our artifact provides the descriptions generated at all three abstraction levels for the same example test~\cite{artifact_abstraction}.} 

Several properties remain invariant across levels to ensure consistency and realism. All descriptions are written as single continuous paragraphs in natural language, without lists, markup, or code blocks. Each description is self-contained, allowing the test to be replicated without assuming pre-existing test suite dependencies; it leaves test names unspecified and concludes by listing the testing framework, assertion library, and mocking library for reproducibility. \change{Finally, the descriptions are structurally neutral: they narrate the test's behavior in plain execution order and do not mark setup or teardown boundaries, use BDD-style phrasing, or otherwise hint at a particular test generation strategy, ensuring that no evaluated approach is favored by the input's structure.}

\subsection{Dataset Statistics}
\begin{wraptable}{r}{0.5685\textwidth}
\vspace{-1em}
\caption{Overview of the \name evaluation dataset.}
\centering
\footnotesize
\setlength{\tabcolsep}{3pt}        
\renewcommand{\arraystretch}{0.9} 
\begin{tabular}{lrrrrrr}
\toprule
& \textbf{Tot} & \textbf{Avg} & \textbf{P25} & \textbf{P50} & \textbf{P75} & \textbf{P90} \\
\midrule
Projects & 20 & -- & -- & -- & -- & -- \\
Classes & 4,482 & 224 & 68 & 141 & 346 & 452 \\
Methods & 33,712 & 1,686 & 600 & 1,349 & 2,626 & 3,531 \\
NCLOC & 246,970 & 12,349 & 4,002 & 9,234 & 16,907 & 25,470 \\
\midrule
Ground-truth tests & 488 & 24 & 4 & 12 & 25 & 48 \\
Test desc. & 1,464 & 73 & 12 & 35 & 76 & 143 \\
\midrule
Focal cls./test & 768 & 1.57 & 1 & 1 & 2 & 3 \\
Focal mtd./test & 1,457 & 2.99 & 1 & 2 & 4 & 6 \\
\bottomrule
\end{tabular}
\label{tab:dataset}
\end{wraptable}

Table~\ref{tab:dataset} summarizes the evaluation dataset. We set a cutoff date of January~31,~2025---the latest knowledge cutoff among evaluated models (\S\ref{subsec:setup})---and retained only tests committed after this date. Of the initial 25 projects, 20 contained tests meeting this criterion, yielding 488 ground-truth tests and 1,464 descriptions across three abstraction levels. 

\subsection{Dataset Validation}
\label{subsec:dataset_val}

\change{Given that LLMs were used to construct the dataset, we assess its quality through a systematic manual analysis of a sample of the generated descriptions. We randomly select 50 entries from each abstraction level, for a total of 150 entries (10.25\% of the dataset), and have one author annotate them in randomized, mixed order along two dimensions: \emph{abstraction fit}, whether a description conforms to its intended abstraction level, and \emph{fidelity}, whether it accurately represents the underlying test.}

\change{For abstraction fit, the annotator is shown the description, the test context (\S\ref{subsec:context_ext}), and the abstraction level definitions (\S\ref{subsec:abstraction_desc}), and judges which level it best matches on a five-point ordinal spectrum---\emph{low}, \emph{low-medium}, \emph{medium}, \emph{medium-high}, and \emph{high}---whose intermediate points capture blends of adjacent levels. Mapping these points to integers (low $=1$ through high $=5$), the three dataset levels sit two points apart, with each intermediate point a half-step between them. For fidelity, the annotator rates whether the description accurately represents the test and characterizes the behavior under test on a four-point Likert scale (\emph{strongly disagree} to \emph{strongly agree}); we omit a neutral option because test correctness admits no meaningful middle ground.}

\change{To establish the reliability of single-annotator labeling, three authors (including the annotator) aligned on the criteria and then independently labeled the same 20 entries. We measure agreement with Krippendorff's alpha (ordinal difference function)~\cite{hayes2007answering} and Gwet's AC2~\cite{gwet2014handbook}, a chance-corrected coefficient stable under skewed rating distributions. For abstraction fit, the authors obtained an alpha of 0.949 and an AC2 of 0.930, indicating near-perfect agreement. For fidelity, they obtained an alpha of $-0.022$ and an AC2 of 0.901. The negative alpha reflects a known paradox of chance-corrected coefficients~\cite{quarfoot2016robust}: because nearly all sampled descriptions were indeed faithful and rated \emph{strongly agree}, estimated chance agreement is so high that alpha collapses even when raters agree on almost all items. AC2, robust to this degeneracy, is the more representative estimate. These results support the reliability of the single annotator's labels on the remaining entries.}

\change{We then compare the annotator's labels on the full sample against the intended labels to determine whether the LLM adhered to the abstraction definitions and faithfully represented the tests. For abstraction fit, we report the exact-match rate, the within-half-step match rate, the mean absolute error (MAE), and the signed bias, all computed over the integer mapping. The annotator's judgments matched the intended level exactly for 56.7\% of entries and within a half-step for 95.3\%, with an MAE of 0.48 and a signed bias of $-0.48$. The high within-half-step rate indicates that the descriptions largely realize their intended abstraction levels; however, the signed bias equaling the MAE in magnitude reveals that every deviation shifts downward, i.e., the LLM tends to include more specific detail than the higher abstraction levels prescribe (\S\ref{sec:threats}). For fidelity, which has no ground truth, the mean rating of 3.94 on the four-point scale indicates that the descriptions almost always precisely capture the underlying test. Overall, the synthetic dataset largely conforms to our abstraction definitions and faithfully represents the ground-truth tests.}
\section{Evaluation}
\label{sec:evaluation}

\subsection{Research Questions}

Our evaluation of \name focuses on the following research questions:

\begin{itemize}[leftmargin=*]

    \item \textbf{RQ1 (Compilability):} What proportion of tests generated by \name are syntactically valid and successfully compile?
    \item \textbf{RQ2 (Coverage):} To what extent do \name-generated tests replicate the code coverage achieved by their corresponding developer-written tests?
    \item \textbf{RQ3 (Structural Fidelity):} How well do \name-generated tests reproduce the structural characteristics of their corresponding developer-written tests?
    \item \textbf{RQ4 (Abstraction Sensitivity):} How does the level of abstraction in the NL test descriptions affect the coverage overlap and structural fidelity of generated tests?
    \item \textbf{RQ5 (Complexity Sensitivity):} How does the complexity of the target test, measured as the number of focal methods it exercises, affect \name's generation quality?
    \item \textbf{RQ6 (Tool Usage):} How does the \name agent allocate tool calls across functional categories, and what patterns emerge in its tool invocation sequences?
    \item \textbf{RQ7 (Cost):} What is the cost of generating tests with \name compared to baseline approaches?
\end{itemize}

\subsection{Experiment Setup}
\label{subsec:setup}

\paragraph{Baseline Tool.} 
We use the official Gemini CLI agent as our baseline, given its demonstrated capability on complex SE tasks~\cite{merrill2026terminal, fu2025automatically}. \change{To ensure a fair comparison and mitigate bias from underspecified prompts, we give the agent a detailed user prompt with structured guidance on localization, dependency identification, and test placement, construction, execution, and verification; we retain the default system prompt so the agent operates as it would for a typical user. The full prompt is available in our artifact~\cite{artifact}.} Our framework is fully automated: it constructs isolated per-project sandboxes, removes existing test suites, executes each prompt, collects execution-level metrics, and organizes results for downstream analysis. We disable web crawling to prevent data leakage, and for each generated test, we store the code, file path, and corresponding dataset entry. To prevent contamination across generations, we process inputs sequentially, resetting the project to an anchored commit and clearing the test suite between generations.

\textit{\name.} 
We follow a similar setup for \name, with one key difference: since our approach produces an application code index, we perform this indexing step once per project after removing the existing test suite, using Qwen3 Embedding 8B~\cite{zhang2025qwen3} for vector representations. To manage computation costs while ensuring sufficient reasoning capacity, we impose iteration caps such that the average cost per input remains under \$0.75, with a \$6 worst-case ceiling for the most expensive models. Across all models, we enable parallel tool calling and set the temperature to 0.3, following existing agentic approaches~\cite{wang2025maintaincoder, zhang2025empowering}.

\textit{Models.} 
We evaluate four models: Gemini 2.5 Pro (G-2.5 Pro) and Gemini 2.5 Flash (G-2.5 Flash)~\cite{comanici2025gemini} for Gemini CLI compatibility, along with two open-source models of varying sizes---Qwen3-Coder (QC)~\cite{qwen3}, a 480B parameter model with 35B activated parameters, and Devstral Small 2 (DS)~\cite{devstral}, a 24B parameter model. We access open-source models through OpenRouter~\cite{openrouter} and Gemini models through Google Cloud~\cite{google_cloud}.

\textit{Evaluation Protocol.}
Following standard practice in agentic software engineering benchmarks~\cite{huang2023agentcoder, jimenez2023swe, hong2023metagpt}, we report pass@1 results from a single execution run per input. This reflects realistic deployment scenarios where practitioners expect correct output on the first attempt and accounts for the substantial computational costs of LLM-based agentic systems.

\subsection{Evaluation Metrics}
\label{subsec:criteria}

We apply a systematic grader that assesses the quality of each generated test across several key metrics by comparing it against its ground-truth test. To prevent compilation conflicts, we insert each generated test file into the existing test suite and execute tests one at a time.

\textit{Compilability.}
We first assess whether a generated test is syntactically correct by compiling it using the project’s build tools. Because inputs are processed individually and the project is guaranteed to compile in isolation prior to test generation, any compilation errors observed can be directly attributed to the generated test.

\textit{Coverage Overlap.}
We measure coverage overlap, defined as the proportion of application classes, methods, lines, and branches covered by the ground-truth test that are also covered by the generated test. To collect coverage data, we use JaCoCo~\cite{jacoco}, executing the ground-truth test and the generated test separately, and computing the set intersection of their respective coverage profiles. After evaluation, a cleanup phase restores the project to its original state.

\textit{Structural Similarity.}
We measure structural test properties using static analysis to assess whether generated tests replicate key characteristics of the ground truth. We employ JavaParser~\cite{javaparser} together with the Hamster~\cite{pan2025hamster} heuristics to extract four properties from each test: (1) type instantiations, identified via constructor calls; (2) assertion types, categorized using Hamster’s assertion classification module; (3) method calls, tracked by receiver type and method name; and (4) focal methods, identified using Hamster’s focal-method heuristics. For each property, we compute precision (the fraction of generated elements that match the ground truth) and recall (the fraction of ground-truth elements that appear in the generated test), and report the F1 score.

\subsection{Evaluation Results}
\label{subsec:results}
\subsubsection{RQ1: Compilability} 
\label{subsubsec:rq1}

\begin{wrapfigure}{r}{0.56\textwidth}
    \centering
    \includegraphics[width=\linewidth]{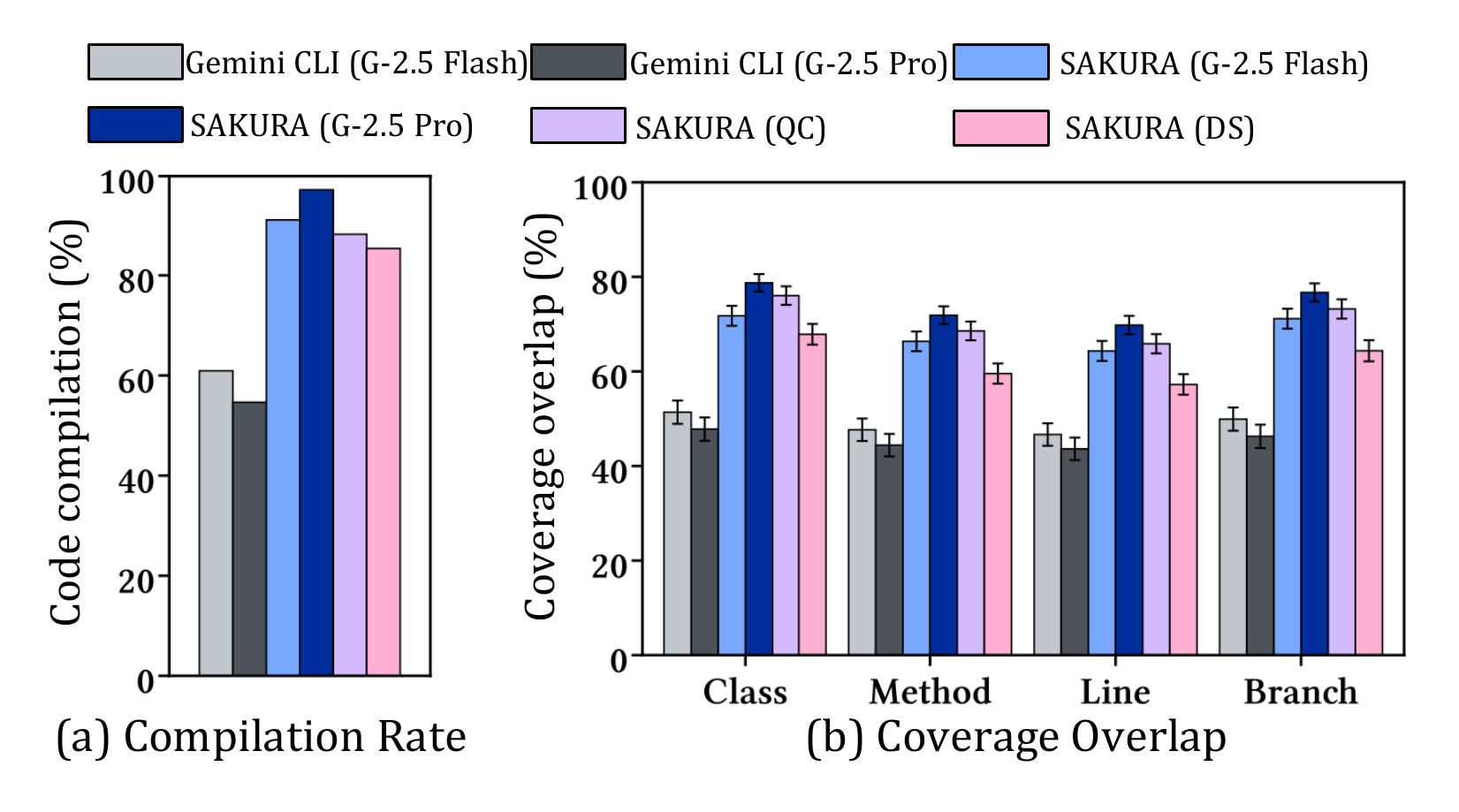}
    \caption{(a) Compilation rate and (b) coverage overlap comparison.}
    \label{fig:rq1-2}
    \vspace{-8pt}
\end{wrapfigure}

We first evaluate the effectiveness of \name and baselines in generating compilable test cases. As shown in Figure~\ref{fig:rq1-2}(a), Gemini CLI consistently struggles (60.9\% with Flash, 54.6\% with Pro), while \name achieves substantially higher rates across all models (Pro: 97.3\%, Flash: 91.3\%, Qwen3-Coder: 88.4\%, Devstral: 85.4\%). Notably, \name with open-source models exceeds Gemini CLI with proprietary models---a pattern that persists across all subsequent RQs.

Our analysis of the failing tests reveals that Gemini CLI's single-agent design and limited toolset lead it to generate tests without sufficient validation, frequently omitting imports, invoking inaccessible methods, or referencing nonexistent classes. \name mitigates these issues through explicit localization and modeling that grounds NL steps in concrete code before synthesis, combined with iterative compile-execute feedback.

\textit{Implication.} Compilability is a prerequisite for any downstream utility---tests that fail to compile cannot be executed, measured, or integrated into CI. These results highlight the importance of aligning tool design with problem structure: smaller models that perform poorly in conventional agentic workflows become competitive when embedded in \name's tool-rich, task-decomposed architecture. Closing the remaining gap via fine-tuning or model-merging~\cite{finetuning1, finetuning2, merging} is a promising future direction.

\begin{findingbox}
\name achieves substantially higher compilability than Gemini CLI, with improvements of 50--78\% (30--43 percentage points) depending on the underlying model. Even \name with smaller models like Devstral Small 2 (85.4\%) outperforms Gemini CLI using Gemini 2.5 Pro (54.6\%) by 56.3\% (30.7 percentage points).
\end{findingbox}

\subsubsection{RQ2: Coverage}
\label{subsubsec:rq2}

We next evaluate how well generated tests exercise the same application behavior as the ground truth using coverage overlap. As shown in Figure~\ref{fig:rq1-2}(b), Gemini CLI with Gemini 2.5 Pro achieves 44--48\% coverage overlap across metrics, while \name with the same model reaches 70--79\%---a 26--31 percentage point (pp) improvement. Even open-source models under \name exceed the baseline's proprietary configurations: Qwen3-Coder achieves 73.2\% branch coverage versus 46.3\% for Gemini CLI with Gemini 2.5 Pro.

These gains stem from \name's decomposition-based workflow, which maps each described action to concrete application logic before generation. This reduces the risk of omitting essential behavior---particularly branching conditions and control flow implied in the description---that end-to-end approaches often miss.

\textit{Implication.} Coverage overlap reflects how accurately an approach captures the semantic intent of test descriptions. End-to-end agents must simultaneously interpret descriptions, localize code, and generate tests, increasing omission likelihood. \name's staged approach distributes this complexity across specialized agents, producing tests that better reflect the full behavioral intent.

\begin{findingbox}
With the same models, \name achieves 38--66\% (18--31 percentage points) higher coverage overlap than Gemini CLI. \name with open-source models (73\% branch coverage for Qwen3-Coder, 64\% for Devstral Small 2) outperforms Gemini CLI with Gemini 2.5 Pro (46\%), demonstrating that agentic architecture can compensate for model capacity.
\end{findingbox}

\subsubsection{RQ3: Structural Fidelity}
\label{subsubsec:rq3}

\begin{figure}[t]
    \centering
    \includegraphics[width=0.96\linewidth]{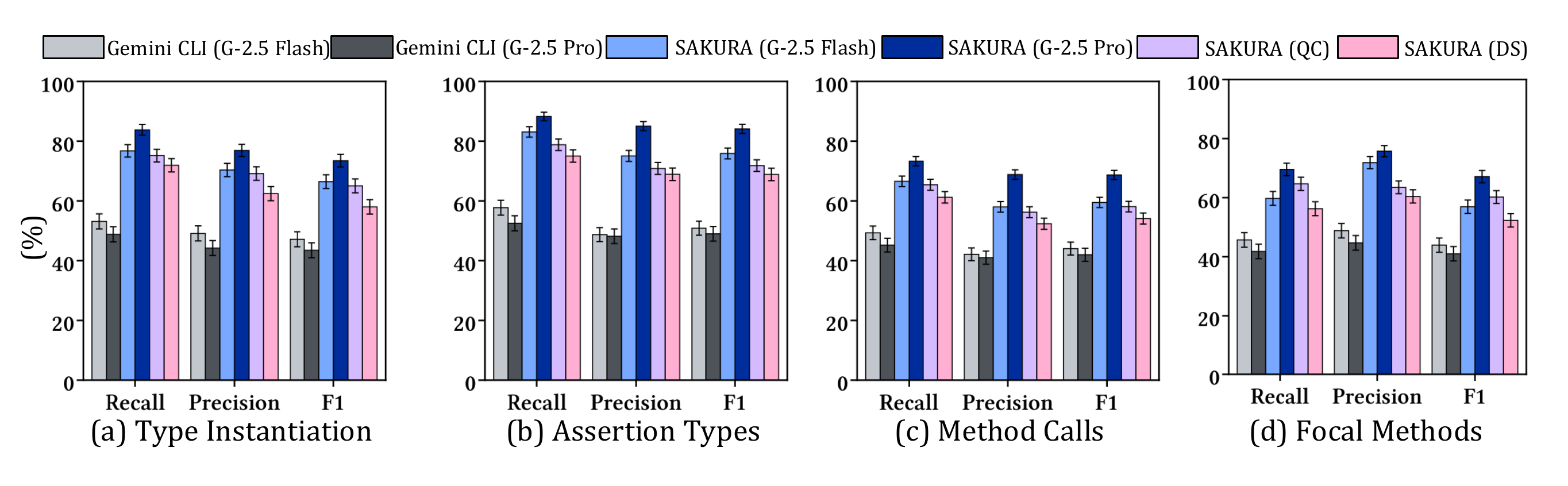}
    \caption{Structural fidelity metrics.} 
    \label{fig:rq3}
\end{figure}

While coverage overlap captures dynamic behavior, it does not fully characterize structural similarity between generated and ground-truth tests. We analyze four static properties---instantiated types, assertion types, invoked methods, and focal methods---computing precision, recall, and F1. We prioritize recall, as lower precision may reflect beneficial additions such as extra setup or cleanup behavior. Figure~\ref{fig:rq3} summarizes the results. Gemini CLI achieves 42--58\% recall across properties, ranging from 42--46\% for focal methods to 52--58\% for assertion types. \name improves recall by 14--36 pp with the same models: Gemini 2.5 Pro achieves 83.8\% for instantiated types, 88.3\% for assertions, 73.3\% for method calls, and 69.6\% for focal methods. Open-source models under \name outperform the proprietary baseline by 14--26 pp, with Qwen3-Coder reaching 65--79\% recall.

The largest gains appear on the two properties most closely tied to test intent: high assertion-type recall indicates \name recovers developers' verification logic, while improved focal-method recall shows it exercises intended behavior rather than peripheral functionality---consistent with \name's localization strategy, which grounds each BDD step in concrete application elements.

\textit{Implication.} Structural fidelity determines whether generated tests integrate naturally into existing test suites. \name produces tests that both execute correct code paths (\S~\ref{subsubsec:rq2}) and structurally resemble developer-written tests---an important distinction for maintainability, as tests with familiar structure are easier to understand, modify, and trust.

\begin{findingbox}
\name improves structural recall by 31--72\% (14--36 percentage points) over Gemini CLI with the same models, with consistent gains across all four properties, including focal method identification (70\% vs. 42\% using Gemini~2.5~Pro). Open-source models under \name still surpass the proprietary baseline, with Qwen3-Coder reaching 65--79\% recall.
\end{findingbox}

\subsubsection{RQ4: Abstraction Sensitivity}
\label{subsubsec:rq4}

\begin{figure}[t]
    \centering
    \includegraphics[width=0.96\linewidth]{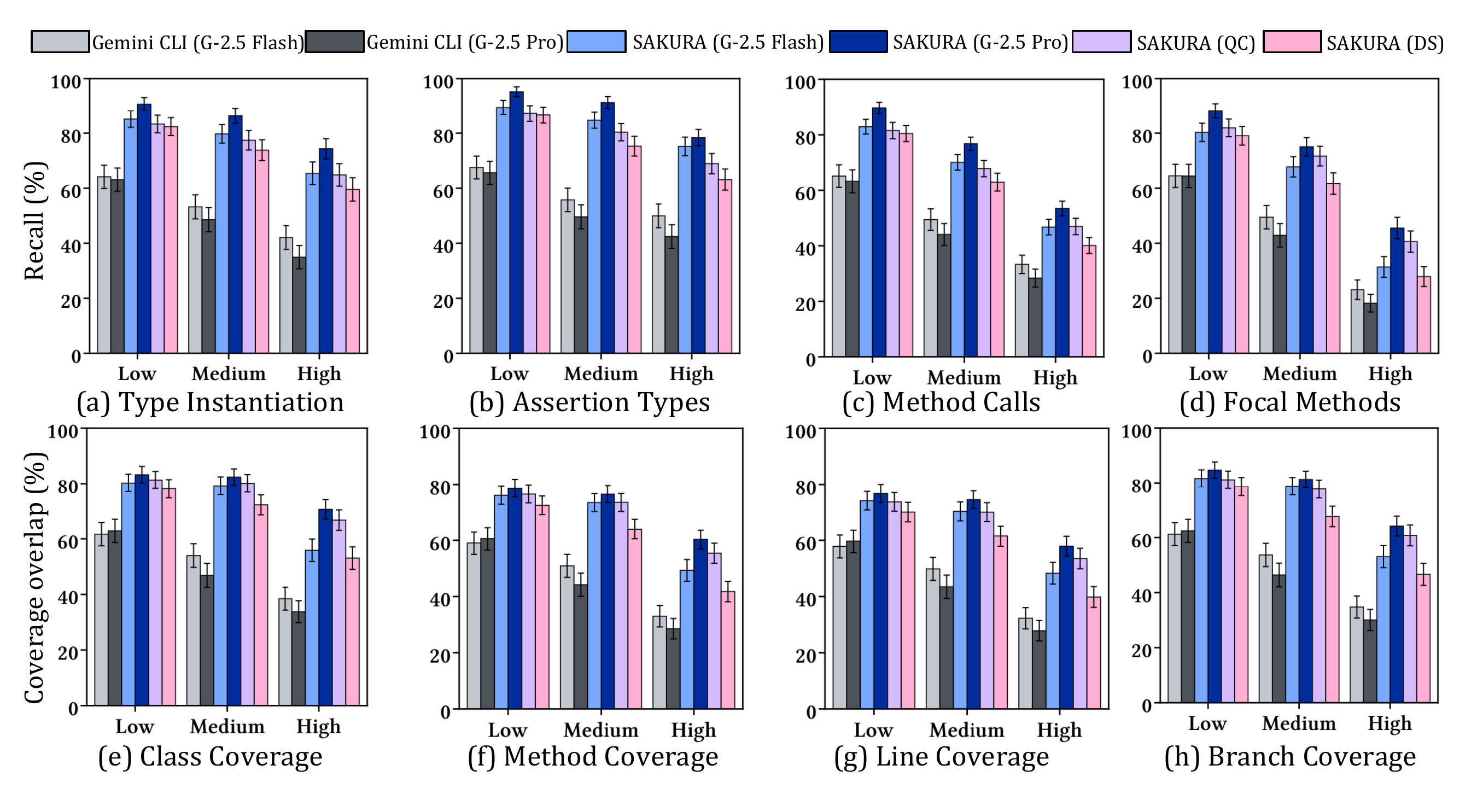}
    \caption{Abstraction sensitivity metrics.} 
    \label{fig:rq4}
\end{figure}

In real-world use, test descriptions vary widely in specificity. To evaluate robustness to this variation, we stratify results by abstraction level (low, medium, high) and report structural recall alongside coverage overlap. Figure~\ref{fig:rq4} depicts our findings.

All approaches degrade as abstraction increases, but \name maintains a substantial advantage at every level. At low abstraction, Gemini CLI achieves 63--68\% structural recall while \name with Gemini 2.5 Pro reaches 88--95\%. At high abstraction, this gap widens: Gemini CLI's focal-method recall drops to 18--23\%, whereas \name with Gemini 2.5 Pro retains 45.5\%. For focal-method recall at high abstraction, open-source models under \name still outperform the proprietary baseline (Qwen3-Coder: 40.6\% vs. Gemini CLI with Pro: 18.2\%). Coverage overlap trends are similar, with \name maintaining 58--71\% overlap at high abstraction versus 28--38\% for the baseline. Among structural properties, focal methods degrade most sharply while assertion types remain robust, suggesting verification intent is more resilient to abstraction than behavioral localization.

\textit{Implication.} Robustness to abstraction determines real-world applicability, since users vary in codebase familiarity and will naturally write descriptions at different specificity levels. These findings indicate that \name's decomposition and localization phases provide increasing value as input specificity decreases. By explicitly grounding abstract steps in concrete application elements before test construction, \name compensates for the missing implementation details that cause single-agent approaches to diverge from intended behavior. This architectural separation---where the localization agent resolves ambiguity independently of code generation---allows \name to bridge the semantic gap that widens with abstraction, rather than forcing a single agent to simultaneously interpret vague intent and produce correct code. 

Nevertheless, performance at high abstraction remains substantially below that of low-abstraction inputs, suggesting room for improvement. Future work could explore uncertainty-aware localization, where the agent explicitly quantifies confidence in its candidate mappings and triggers alternative strategies---such as parallelized multi-hypothesis generation or interactive clarification---when ambiguity is high. 

\begin{findingbox}
\name is substantially more robust to input abstraction: at high abstraction, \name with Gemini~2.5~Pro achieves 150.0\% (27.3 pp) higher focal-method recall and 110.5\% (33.2 pp) higher average coverage overlap than Gemini CLI with the same model, and open-source models under \name still outperform the proprietary baseline.
\end{findingbox}

\subsubsection{RQ5: Complexity Sensitivity}
\label{subsubsec:rq5}

\begin{figure}[t]
    \centering
    \includegraphics[width=0.96\linewidth]{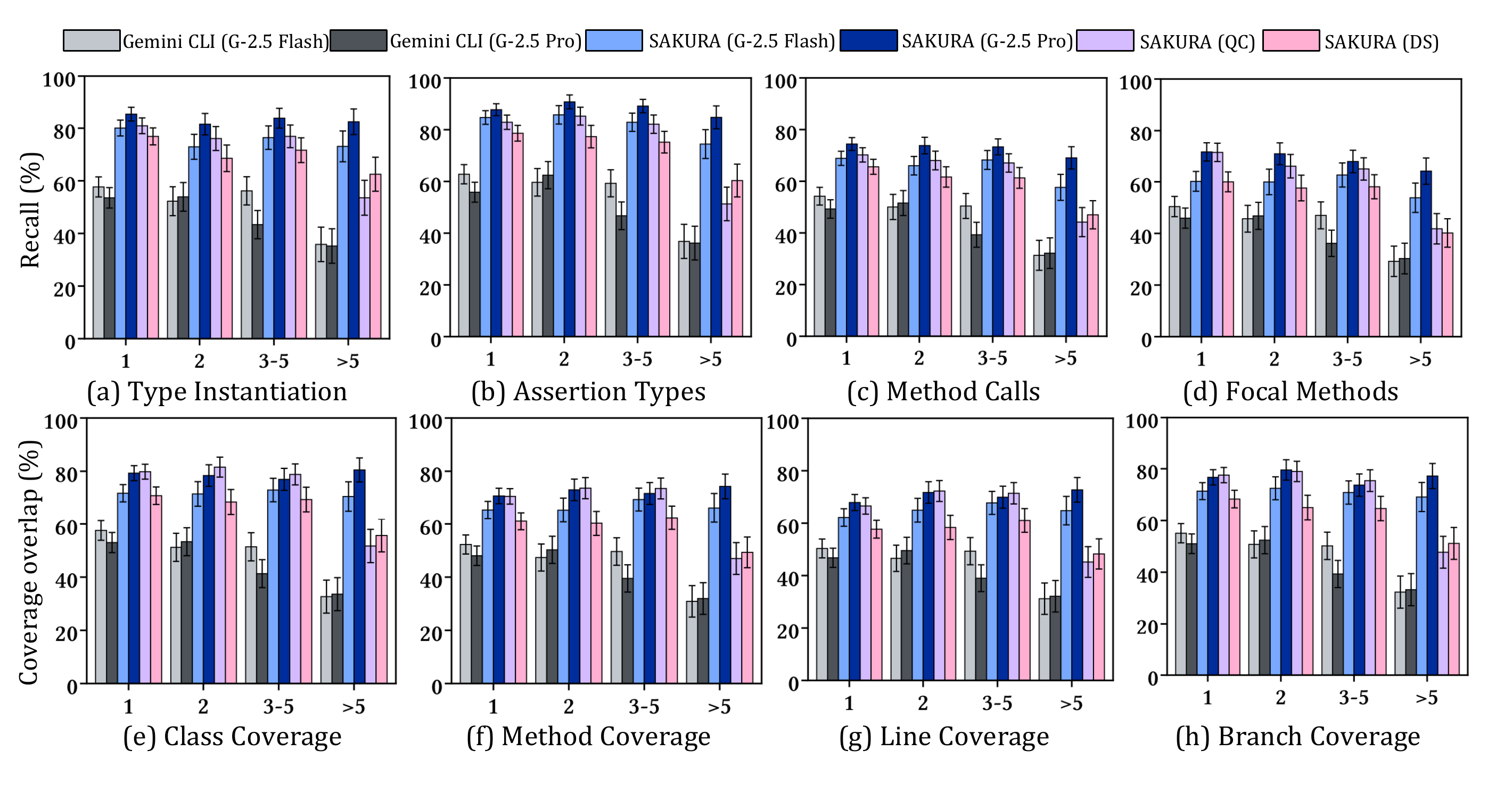}
    \caption{Complexity sensitivity.} 
    \label{fig:rq5}
\end{figure}

While RQ4 (\S~\ref{subsubsec:rq4}) evaluated \name's ability to handle varying levels of detail in test descriptions, we now assess \name's capacity to generate structurally complex tests (\S\ref{sec:intro}), using focal-method count as a proxy for complexity. We group results into four buckets (1, 2, 3--5, >5 focal methods) to ensure balanced distribution. Figure~\ref{fig:rq5} presents the results.

Gemini CLI degrades sharply with complexity, while \name remains stable: the baseline falls from 46--63\% across metrics on single-focal tests to 29--37\% at >5 focal methods, whereas \name with Gemini 2.5 Pro achieves 68--88\% and sustains 64--85\%, respectively. The gap thus widens with complexity, reaching 25--49~pp at >5 focal methods with the same model. Open-source models under \name still exceed the proprietary baseline at high complexity (Qwen3-Coder: 42--54\% vs.\ Gemini CLI Pro: 30--36\%), though their steeper degradation (21--32~pp for Qwen3-Coder) suggests smaller models struggle with the richer context that decomposition produces for complex inputs.

These results reflect \name's architectural advantages for complex test generation. When tests involve multiple focal methods, single-agent approaches must simultaneously reason about diverse behaviors and coordinate across components---a combinatorial challenge that compounds quickly. \name's decomposition model and localization agent ground each focal method independently, transforming complex multi-focal tests into tractable subtasks.

\textit{Implication.} 
Resilience to complexity is essential for practical utility: integration and system tests frequently exercise multiple focal methods. The widening gap (34--49 pp at high vs. 21--32 pp at low complexity with Gemini~2.5~Pro) indicates \name's benefits compound with task difficulty---structured decomposition is necessary for non-trivial test generation. However, the capacity-complexity tradeoff with open-source models suggests future work on adaptive decomposition strategies modulating context density based on model capacity and detected test complexity.

\begin{findingbox}
As focal-method count increases from 1 to >5, Gemini CLI declines by about 20 pp while \name with Gemini~2.5~Pro sustains 64--85\% across metrics, widening the same-model gap from 19--59\% (10--32 pp) to 84--140\% (25--49 pp). Open-source models under \name still surpass the proprietary baseline at high complexity, albeit with steeper degradation.
\end{findingbox}

\subsubsection{RQ6: Tool Usage}
\label{subsubsec:rq6}
A major component of an agent's success is its tool-calling behavior, which results from its curated tool set and how those tools are prompted in the agent design. We analyze how agents allocate tool calls across five functional categories: retrieval, inspection, generation, validation, and orchestration. Table~\ref{tab:tool-categories} summarizes this taxonomy, and Figure~\ref{fig:rq6} depicts our results.

\begin{table}[t]
    \centering
    \footnotesize
    \setlength{\tabcolsep}{4pt}
    \renewcommand{\arraystretch}{1.05}
    \caption{Tool-usage categorization by functional role.}
    \label{tab:tool-categories}
    \begin{tabular}{@{}l l l l@{}}
        \toprule
        \textbf{Category} & \textbf{Description} & \textbf{\name} & \textbf{Gemini CLI} \\
        \midrule
        Retrieval & Semantic/pattern search & \texttt{method\_vector\_search}, \texttt{class\_vector\_search}, & \texttt{glob}, \\
                  &                         & \texttt{reachable\_methods\_search} & \texttt{search\_file\_content} \\
        \midrule
        Inspection & Static code reading & \texttt{extract\_method\_source}, \texttt{get\_method\_details}, & \texttt{read\_file}, \\
                   &                     & \texttt{get\_call\_site\_details}, \ldots & \texttt{list\_directory} \\
        \midrule
        Generation & Code writing & \texttt{generate\_test\_code}, & \texttt{write\_file}, \texttt{replace} \\
                   &              & \texttt{modify\_step\_comment} & \\
        \midrule
        Validation & Compile/execute & \texttt{compile\_and\_execute\_test} & \texttt{run\_shell\_command} \\
        \midrule
        Orchestration & Agent coordination & \texttt{call\_localization\_agent}, & \texttt{delegate\_to\_agent} \\
                      &                    & \texttt{call\_composition\_agent} & \\
        \bottomrule
    \end{tabular}
\end{table}

Gemini CLI exhibits highly skewed distributions that vary dramatically between models: Flash dedicates 69.7\% to inspection with only 7.2\% to generation, while Pro inverts this pattern (49.7\% generation, 29.4\% inspection). In contrast, \name produces consistently balanced distributions across all models: retrieval (23--35\%), inspection (28--45\%), generation (7--15\%), validation (7--11\%), and orchestration (14--20\%). This balance holds even for open-source models. 

\begin{table}[t]
\centering
\footnotesize
\begin{minipage}[c]{0.4\textwidth}
    \centering
    \includegraphics[width=\linewidth]{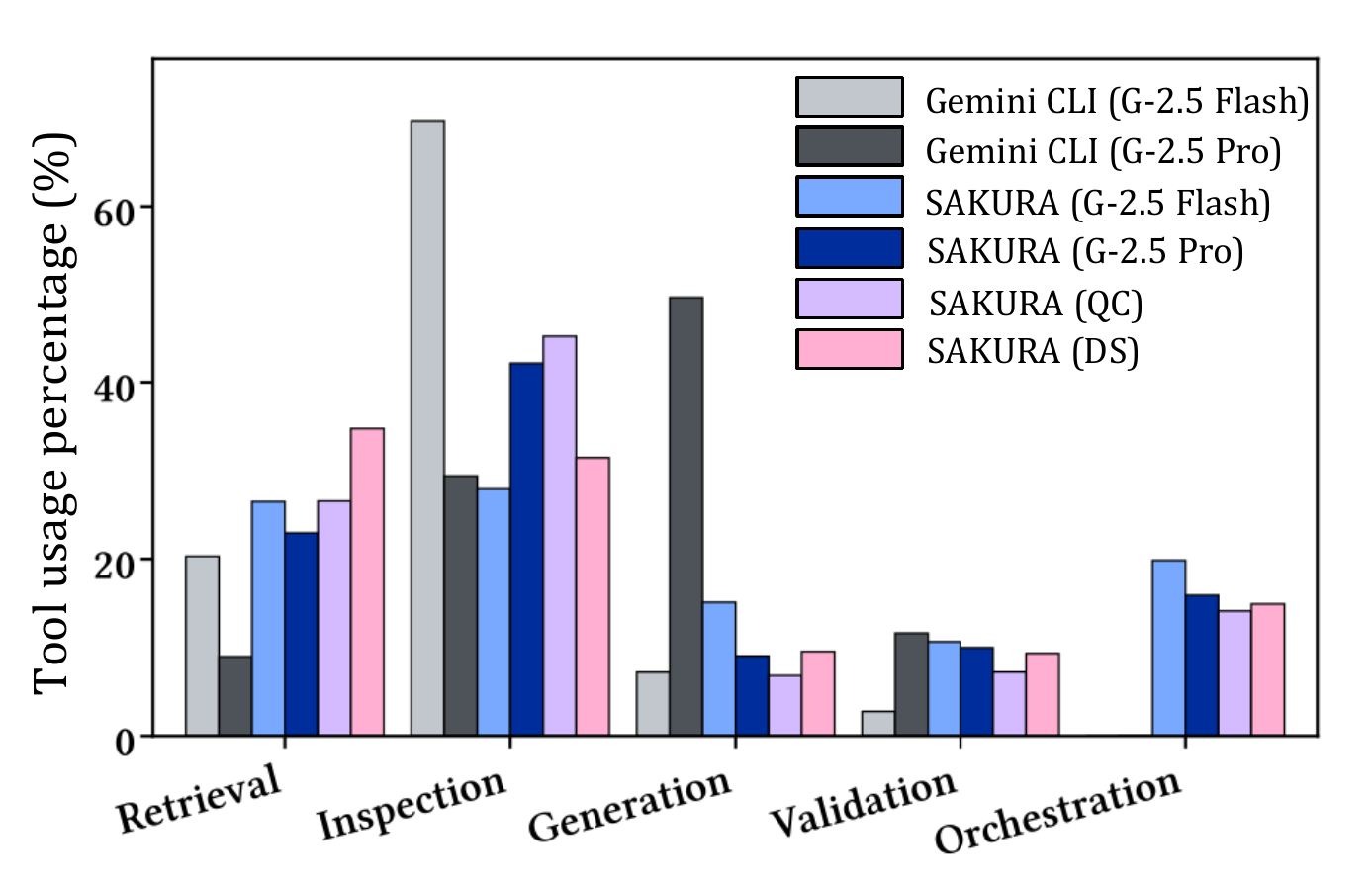}
    \captionof{figure}{Tool usage distribution by category.}
    \label{fig:rq6}
\end{minipage}%
\hspace{0.04\textwidth}%
\begin{minipage}[c]{0.35\textwidth}
    \centering
    \caption{Average cost per input.}
    \label{tab:avg-cost}
    \scriptsize
    \renewcommand{\arraystretch}{1.2}
    \setlength{\tabcolsep}{4pt}
    \begin{tabular}{lc}
        \toprule
        Approach & Cost (\$) \\
        \midrule
        Gemini CLI (G-2.5 Flash)    & 0.02 \\
        Gemini CLI (G-2.5 Pro)      & 0.13 \\
        \name (G-2.5 Flash) & 0.10 \\
        \name (G-2.5 Pro)  & 0.52 \\
        \name (Qwen3-Coder)        & 0.07 \\
        \name (Devstral Small 2)     & 0.02 \\
        \bottomrule
    \end{tabular}
\end{minipage}
\end{table}

These patterns reflect architectural differences. Gemini CLI's single-agent and broad tool design requires autonomous, unguided decisions about when to explore, generate, and validate---leading to model-specific biases. \name's multi-agent architecture enforces consistent workflows through explicit phase separation and focused tool construction.

\textit{Implication.} Tool allocation patterns explain the performance differences observed in prior RQs: erratic distributions indicate unprincipled workflows that skip essential phases. While Gemini CLI's allocation shifts by up to 43 percentage points between models, \name maintains distributions within 17 percentage points across configurations. This suggests well-designed scaffolding can compensate for model-level planning differences, enabling smaller models to follow principled workflows that larger models might discover independently. The orchestration overhead (14--20\%) represents \name's investment in coordination that single-agent approaches neglect.

\begin{findingbox}
\name maintains balanced tool usage across all models, with all five categories consistently represented and cross-model variation within 17 pp, whereas Gemini CLI's allocation is highly skewed and shifts by up to 43 pp between models. \name's domain-specific tools enable open-source models to achieve workflows comparable to proprietary ones.
\end{findingbox}

\subsubsection{RQ7: Cost}
\label{subsubsec:rq7}

For widespread adoption, LLM-based techniques must be cost-efficient. Table~\ref{tab:avg-cost} reports the average cost per generated test across approaches.

\name incurs higher costs than Gemini CLI when using the same underlying model, reflecting the additional inference required for decomposition, localization, and iterative refinement. With Gemini 2.5 Pro, \name costs approximately 4 times more than the baseline. However, \name with open-source models offers a compelling trade-off: at \$0.07 per test, Qwen3-Coder costs roughly half of Gemini CLI with Gemini 2.5 Pro while substantially outperforming it across all metrics (\S~\ref{subsubsec:rq1}--\ref{subsubsec:rq5}). Devstral Small 2 matches Gemini CLI's Flash-tier pricing at \$0.02 per test yet still exceeds the baseline's performance. These results demonstrate that \name's structured workflow enables smaller, cheaper models to perform well.

\textit{Implication.} The cost analysis highlights a clear deployment trade-off. When cost is not a constraint, \name with Gemini~2.5~Pro achieves the highest test quality. In cost-sensitive settings, \name paired with open-source models offers substantial improvements over the baseline at comparable or lower cost. Future work includes adaptive model selection, where expensive models are invoked selectively for high-abstraction or complex inputs, while simpler cases are handled by lightweight models. Additional directions include speculative generation---using inexpensive models to produce multiple candidates, then a single costly call to select or merge them---and localization caching to amortize grounding costs across tests within the same project.

\begin{findingbox}
\name with open-source models (Qwen3-Coder, Devstral Small 2) achieves greater performance than Gemini CLI with proprietary models at equal or lower cost, demonstrating that structured workflows can offset model capacity and enable cost-effective deployment.
\end{findingbox}

\section{Related Work}
\label{sec:related}

NL-to-test generation poses two key challenges: constructing paired NL-test datasets and synthesizing compilable, executable tests from descriptions. We review existing work and highlight limitations that motivate our approach.

\textit{Benchmarks and Datasets.}
Benchmarks such as SWE-bench~\cite{jimenez2023swe}, HumanEval~\cite{chen2021evaluating}, and MBPP~\cite{austin2021program} evaluate code generation from NL using test cases as oracles---valuable for general code synthesis but not focused on test generation specifically. TDD-bench Verified~\cite{ahmed2024tdd} extends this paradigm by generating tests from issue reports and verifying fail-then-pass behavior across repository snapshots. However, \name addresses test generation more holistically rather than focusing solely on bug-reproducing tests, and targets Java rather than Python.

Other approaches benchmark NL-to-test generation by pairing descriptions with ground-truth tests, deriving the descriptions from focal-method docstrings in test-to-focal-method datasets such as Methods2Test~\cite{alagarsamy2025enhancing, lops2025system}. This heuristic is brittle: docstrings are often incomplete or misleading, and method names frequently fail to capture the full behavior under test~\cite{wen2019large, tan2012tcomment}. Moreover, these datasets typically contain only unit tests, limiting the behaviors that can be evaluated, and ignore LLM training cutoffs, risking data contamination. We address these concerns with a curation pipeline that uses static analysis to extract sufficient context for semantically aligned descriptions, supports well-defined abstractions for varied description qualities, and filters by LLM cutoff dates.

\textit{Test Generation Approaches.}
Most prior techniques start from code and produce regression-style tests capturing its current behavior. Search-based tools (e.g., EvoSuite~\cite{fraser2011evosuite}, Randoop~\cite{pacheco2007randoop}) explore the implementation to generate tests, and LLM-based approaches (e.g., ChatUniTest~\cite{chen_chatunitest_2024}, TestPilot~\cite{schafer2023empirical}) prompt models with a focal method. \change{Among these, the closest to \name are \textsc{ASTER}~\cite{pan2025aster} and TestForge~\cite{jain_testforge_2025}: \name shares ASTER's use of static analysis and TestForge's execution-feedback loop. Both, however, assume a pre-identified focal method or file for which they generate unit regression tests, whereas \name receives only an unlocalized NL description and must first localize the intended behavior, which may span multiple classes and call sequences.}

A complementary line of work instead takes natural language as its starting point, but relies on stronger assumptions about the input or imposes weaker requirements on the output. One prominent direction generates bug-reproducing tests from issue reports~\cite{nashid2025issue2test, kang2022large, kitsios2025automated,
ahmedotter}; as Plein et al.~\cite{plein2023automatic} observe, however, these approaches depend on well-structured reports whose details effectively supply a localized target, limiting their generalization to vague descriptions of application behavior. Requirements-based techniques similarly presuppose implementation knowledge beyond the requirements themselves, such as focal methods or localized code context~\cite{alagarsamy2025enhancing, ferreira2025acceptance}---even when requirements are mediated through BDD scenarios~\cite{ferreira2025acceptance}---or else produce high-level test descriptions rather than executable tests that integrate into an existing suite~\cite{masuda2025generating, korraprolu2025requirements}.
\change{General-purpose agentic systems (e.g., SWE-agent~\cite{yang2024swe}, CodeAgent~\cite{zhang2024codeagent}, Gemini CLI~\cite{gemini_cli}) dispense with these assumptions altogether---they accept NL input and operate autonomously over the codebase---but lack test-specific abstractions such as static-analysis-based localization, structured execution feedback, and test-aware decomposition; our evaluation quantifies this gap against Gemini CLI directly (\S~\ref{sec:evaluation}).}
\section{Threats to Validity}
\label{sec:threats}

\textit{\indent Internal Validity.}
A primary internal threat is data contamination. We mitigate this risk through several measures: retaining only tests committed after a conservative cutoff date (the latest training cutoff among evaluated models), disabling web access during all runs, and removing the existing test suite before each generation. \change{Another internal threat is the manual dataset validation described in \S\ref{subsec:dataset_val}, which may involve subjectivity and bias. To mitigate this threat, three authors independently labeled a shared subset and achieved near-perfect inter-rater agreement.}

\textit{External Validity.}
Our evaluation focuses on Java projects from Apache Commons, limiting direct generalization to other ecosystems. However, these projects span diverse domains representative of real-world codebases. \change{Moreover, \name's architecture is language-agnostic by design; for instance, the decomposition model (\S\ref{subsec:decomposeNL}) reflects the setup--scenario--teardown structure common to tests across languages and frameworks and contains no Java-specific constructs.} A related concern is that our NL inputs are systematically generated from developer tests rather than written by practitioners. We address this by varying abstraction levels and LLM temperatures, increasing the diversity and realism of the resulting test descriptions.

\textit{Construct Validity.}
We use compilability, coverage overlap, and structural similarity as proxies for test quality. While indirect, these measures capture executability and alignment with developer-written tests, consistent with established quality indicators~\cite{pan2025hamster}. To ensure fairness, we apply the same analysis pipeline uniformly across all approaches. \change{A further construct threat is that our LLM-generated descriptions may not perfectly operationalize the intended abstraction levels. Our validation (\S\ref{subsec:dataset_val}) shows that most descriptions realize their intended levels, and deviations, when they occur, tend to retain more implementation detail than prescribed. Consequently, results at higher abstraction levels may overestimate performance on truly business-level inputs. Note that this bias does not hinder our comparison: since the gap between \name and the baseline grows with abstraction, truly business-level inputs would likely only widen it.}

\textit{Conclusion (Statistical) Validity.}
Following standard agentic-benchmark practice~\cite{jimenez2023swe, huang2023agentcoder}, we report pass@1 metrics from a single generation run per input due to cost constraints. Because LLM agents can exhibit nondeterministic behavior, single-run estimates may vary across reruns. Nevertheless, the scale of our dataset supports stable aggregate estimates.

\section{Summary and Future Work}
\label{sec:summary}

We presented \name, a multi-agent approach for generating structurally complex Java tests from natural language descriptions. \name decomposes each test description into BDD-style structured blocks and processes them through three specialized agents: a localization agent that maps blocks to candidate classes and methods, a composition agent that synthesizes and validates executable test code, and a supervisor agent that coordinates the workflow and verifies correctness.
We evaluated \name on a curated dataset of 1,464 test descriptions paired with developer-written tests from high-quality Java projects. Across four LLMs---two proprietary and two open-source---\name consistently outperformed Gemini CLI in generating compilable tests and achieving higher coverage overlap with developer-written tests. Notably, \name with smaller open-source models surpassed Gemini CLI using larger proprietary models, demonstrating that structured multi-agent coordination can compensate for reduced model capacity.
Although performance with large proprietary models remains higher overall, closing this gap is an important direction for future work. We plan to explore model fine-tuning, model merging, and further task decomposition to improve the effectiveness of smaller models within our framework. Beyond these optimizations, we also aim to extend \name to additional domains, including API test generation and automated test generation from bug reports.
\section{Data Availability}
\label{sec:data}

Our replication package~\cite{artifact} includes the LLM prompts used for \name and Gemini CLI, the evaluation dataset comprising 1,464 test scenarios with corresponding developer-written tests, the pipeline used to construct the \name dataset, and the full multi-agent implementation needed to reproduce our experiments. It also contains the results supporting each RQ, the results of the author-conducted dataset validation, and the outputs for the examples discussed in the paper.

\bibliographystyle{ACM-Reference-Format}
\bibliography{references}

@inproceedings{lops2025system,
  title={A System for Automated Unit Test Generation using Large Language Models and Assessment of Generated Test Suites}, 
  author={Lops, Andrea and Narducci, Fedelucio and Ragone, Azzurra and Trizio, Michelantonio and Bartolini, Claudio},
  booktitle={2025 IEEE International Conference on Software Testing, Verification and Validation Workshops (ICSTW)}, 
  year={2025},
  volume={},
  number={},
  pages={29-36},
  doi={10.1109/ICSTW64639.2025.10962454}
}

@inproceedings{pan2025aster,
    author = {Pan, Rangeet and Kim, Myeongsoo and Krishna, Rahul and Pavuluri, Raju and Sinha, Saurabh},
    title = {ASTER: Natural and Multi-Language Unit Test Generation with LLMs},
    year = {2025},
    publisher = {IEEE Press},
    url = {https://doi.org/10.1109/ICSE-SEIP66354.2025.00042},
    doi = {10.1109/ICSE-SEIP66354.2025.00042},
    booktitle = {2025 IEEE/ACM 47th International Conference on Software Engineering: Software Engineering in Practice (ICSE-SEIP)},
    pages = {413–424},
    numpages = {12},
    location = {Ottawa, ON, Canada}
}

@inproceedings{fraser2011evosuite,
    author = {Fraser, Gordon and Arcuri, Andrea},
    title = {EvoSuite: automatic test suite generation for object-oriented software},
    year = {2011},
    isbn = {9781450304436},
    publisher = {Association for Computing Machinery},
    address = {New York, NY, USA},
    url = {https://doi.org/10.1145/2025113.2025179},
    doi = {10.1145/2025113.2025179},
    booktitle = {Proceedings of the 19th ACM SIGSOFT Symposium and the 13th European Conference on Foundations of Software Engineering},
    pages = {416–419},
    numpages = {4},
    location = {Szeged, Hungary},
    series = {ESEC/FSE '11}
}

@inproceedings{pan2025hamster,
    author = {Pan, Rangeet and Stennett, Tyler and Pavuluri, Raju and Levin, Nate and Orso, Alessandro and Sinha, Saurabh},
    title = {Hamster: A Large-Scale Study and Characterization of Developer-Written Tests},
    year = {2026},
    isbn = {9798400724268},
    publisher = {Association for Computing Machinery},
    address = {New York, NY, USA},
    url = {https://doi.org/10.1145/3786583.3786857},
    doi = {10.1145/3786583.3786857},
    booktitle = {Proceedings of the IEEE/ACM 48th International Conference on Software Engineering: Software Engineering in Practice},
    pages = {106–117},
    numpages = {12},
    location = {
    },
    series = {ICSE-SEIP '26}
}

@misc{javaparser,
  title = {{JavaParser}},
  year = 2025,
  url = {https://github.com/javaparser/javaparser},
  key = {javaparser}
}

@article{alagarsamy2025enhancing,
    author = {Alagarsamy, Saranya and Tantithamthavorn, Chakkrit and Takerngsaksiri, Wannita and Arora, Chetan and Aleti, Aldeida},
    title = {Enhancing large language models for text-to-testcase generation},
    year = {2025},
    issue_date = {Dec 2025},
    publisher = {Elsevier Science Inc.},
    address = {USA},
    volume = {230},
    number = {C},
    issn = {0164-1212},
    url = {https://doi.org/10.1016/j.jss.2025.112531},
    doi = {10.1016/j.jss.2025.112531},
    journal = {J. Syst. Softw.},
    month = dec,
    numpages = {14}
}

@inproceedings{pacheco2007randoop,
    author = {Pacheco, Carlos and Ernst, Michael D.},
    title = {Randoop: feedback-directed random testing for Java},
    year = {2007},
    isbn = {9781595938657},
    publisher = {Association for Computing Machinery},
    address = {New York, NY, USA},
    url = {https://doi.org/10.1145/1297846.1297902},
    doi = {10.1145/1297846.1297902},
    booktitle = {Companion to the 22nd ACM SIGPLAN Conference on Object-Oriented Programming Systems and Applications Companion},
    pages = {815–816},
    numpages = {2},
    location = {Montreal, Quebec, Canada},
    series = {OOPSLA '07}
}

@inproceedings{pacheco2007feedback,
  author={Pacheco, Carlos and Lahiri, Shuvendu K. and Ernst, Michael D. and Ball, Thomas},
  booktitle={29th International Conference on Software Engineering (ICSE'07)}, 
  title={Feedback-Directed Random Test Generation}, 
  year={2007},
  volume={},
  number={},
  pages={75-84},
  doi={10.1109/ICSE.2007.37}
}

@inproceedings{tillmann2008pex,
    author="Tillmann, Nikolai
    and de Halleux, Jonathan",
    editor="Beckert, Bernhard
    and H{\"a}hnle, Reiner",
    title="Pex--White Box Test Generation for .NET",
    booktitle="Tests and Proofs",
    year="2008",
    publisher="Springer Berlin Heidelberg",
    address="Berlin, Heidelberg",
    pages="134--153",
    isbn="978-3-540-79124-9",
    doi = {10.1007/978-3-540-79124-9_10},
}

@inproceedings{lukasczyk2022pynguin,
  author={Lukasczyk, Stephan and Fraser, Gordon},
  booktitle={2022 IEEE/ACM 44th International Conference on Software Engineering: Companion Proceedings (ICSE-Companion)}, 
  title={Pynguin: Automated Unit Test Generation for Python}, 
  year={2022},
  volume={},
  number={},
  pages={168-172},
  doi={10.1145/3510454.3516829}
}

@inproceedings{godefroid2005dart,
    author = {Godefroid, Patrice and Klarlund, Nils and Sen, Koushik},
    title = {DART: directed automated random testing},
    year = {2005},
    isbn = {1595930566},
    publisher = {Association for Computing Machinery},
    address = {New York, NY, USA},
    url = {https://doi.org/10.1145/1065010.1065036},
    doi = {10.1145/1065010.1065036},
    booktitle = {Proceedings of the 2005 ACM SIGPLAN Conference on Programming Language Design and Implementation},
    pages = {213–223},
    numpages = {11},
    location = {Chicago, IL, USA},
    series = {PLDI '05}
}

@inproceedings{lin:2021,
    author = {Lin, Yun and Ong, You Sheng and Sun, Jun and Fraser, Gordon and Dong, Jin Song},
    title = {Graph-based seed object synthesis for search-based unit testing},
    year = {2021},
    isbn = {9781450385626},
    publisher = {Association for Computing Machinery},
    address = {New York, NY, USA},
    url = {https://doi.org/10.1145/3468264.3468619},
    doi = {10.1145/3468264.3468619},
    booktitle = {Proceedings of the 29th ACM Joint Meeting on European Software Engineering Conference and Symposium on the Foundations of Software Engineering},
    pages = {1068–1080},
    numpages = {13},
    location = {Athens, Greece},
    series = {ESEC/FSE 2021}
}

@inproceedings{pasareanu:2010,
  author = {P\u{a}s\u{a}reanu, Corina S. and Rungta, Neha},
  title = {Symbolic PathFinder: Symbolic Execution of Java Bytecode},
  year = {2010},
  doi = {10.1145/1858996.1859035},
  booktitle = {Proceedings of the IEEE/ACM International Conference on Automated Software Engineering},
  pages = {179–180},
}

@article{chen:2010:jss,
  author = {Chen, Tsong Yueh and Kuo, Fei-Ching and Merkel, Robert G. and Tse, T. H.},
  title = {Adaptive Random Testing: The {ART} of Test Case Diversity},
  year = {2010},
  volume = {83},
  number = {1},
  doi = {10.1016/j.jss.2009.02.022},
  journal = {J. Syst. Softw.},
  month = jan,
  pages = {60–66},
}

@inproceedings{ciupa:2008:icse,
    author = {Ciupa, Ilinca and Leitner, Andreas and Oriol, Manuel and Meyer, Bertrand},
    title = {ARTOO: adaptive random testing for object-oriented software},
    year = {2008},
    isbn = {9781605580791},
    publisher = {Association for Computing Machinery},
    address = {New York, NY, USA},
    url = {https://doi.org/10.1145/1368088.1368099},
    doi = {10.1145/1368088.1368099},
    booktitle = {Proceedings of the 30th International Conference on Software Engineering},
    pages = {71–80},
    numpages = {10},
    location = {Leipzig, Germany},
    series = {ICSE '08}
}

@inproceedings{lin:2009:ase,
  author={Lin, Yu and Tang, Xucheng and Chen, Yuting and Zhao, Jianjun},
  booktitle={2009 IEEE/ACM International Conference on Automated Software Engineering}, 
  title={A Divergence-Oriented Approach to Adaptive Random Testing of Java Programs}, 
  year={2009},
  pages={221-232},
  doi={10.1109/ASE.2009.13}
}

@inproceedings{renze2024effect,
    title = "The Effect of Sampling Temperature on Problem Solving in Large Language Models",
    author = "Renze, Matthew",
    editor = "Al-Onaizan, Yaser  and
      Bansal, Mohit  and
      Chen, Yun-Nung",
    booktitle = "Findings of the Association for Computational Linguistics: EMNLP 2024",
    month = nov,
    year = "2024",
    address = "Miami, Florida, USA",
    publisher = "Association for Computational Linguistics",
    url = "https://aclanthology.org/2024.findings-emnlp.432/",
    doi = "10.18653/v1/2024.findings-emnlp.432",
    pages = "7346--7356",
}

@inproceedings{arcuri:2011:issta,
  author = {Arcuri, Andrea and Briand, Lionel},
  title = {Adaptive Random Testing: An Illusion of Effectiveness?},
  year = {2011},
  doi = {10.1145/2001420.2001452},
  booktitle = {Proceedings of the 2011 International Symposium on Software Testing and Analysis},
  pages = {265–275}
}

@inproceedings{plein2023automatic,
    author = {Plein, Laura and Ou\'{e}draogo, Wendk\^{u}uni C. and Klein, Jacques and Bissyand\'{e}, Tegawend\'{e} F.},
    title = {Automatic Generation of Test Cases based on Bug Reports: a Feasibility Study with Large Language Models},
    year = {2024},
    isbn = {9798400705021},
    publisher = {Association for Computing Machinery},
    address = {New York, NY, USA},
    url = {https://doi.org/10.1145/3639478.3643119},
    doi = {10.1145/3639478.3643119},
    booktitle = {Proceedings of the 2024 IEEE/ACM 46th International Conference on Software Engineering: Companion Proceedings},
    pages = {360–361},
    numpages = {2},
    location = {Lisbon, Portugal},
    series = {ICSE-Companion '24}
}

@article{ahmed2024tdd,
  title={TDD-Bench Verified: Can LLMs Generate Tests for Issues Before They Get Resolved?},
  author={Ahmed, Toufique and Hirzel, Martin and Pan, Rangeet and Shinnar, Avraham and Sinha, Saurabh},
  journal={arXiv preprint arXiv:2412.02883},
  year={2024},
  doi={10.48550/arXiv.2412.02883}
}

@article{austin2021program,
  title={Program synthesis with large language models},
  author={Austin, Jacob and Odena, Augustus and Nye, Maxwell and Bosma, Maarten and Michalewski, Henryk and Dohan, David and Jiang, Ellen and Cai, Carrie and Terry, Michael and Le, Quoc and others},
  journal={arXiv preprint arXiv:2108.07732},
  year={2021},
  doi={10.48550/arXiv.2108.07732},
}

@inproceedings{
    jimenez2023swe,
    title={{SWE}-bench: Can Language Models Resolve Real-world Github Issues?},
    author={Carlos E Jimenez and John Yang and Alexander Wettig and Shunyu Yao and Kexin Pei and Ofir Press and Karthik R Narasimhan},
    booktitle={The Twelfth International Conference on Learning Representations},
    year={2024},
    url={https://openreview.net/forum?id=VTF8yNQM66}
}

@article{chen2021evaluating,
  title={Evaluating Large Language Models Trained on Code},
  author={Mark Chen and Jerry Tworek and Heewoo Jun and Qiming Yuan and Henrique Ponde de Oliveira Pinto and Jared Kaplan and Harri Edwards and Yuri Burda and Nicholas Joseph and Greg Brockman and Alex Ray and Raul Puri and Gretchen Krueger and Michael Petrov and Heidy Khlaaf and Girish Sastry and Pamela Mishkin and Brooke Chan and Scott Gray and Nick Ryder and Mikhail Pavlov and Alethea Power and Lukasz Kaiser and Mohammad Bavarian and Clemens Winter and Philippe Tillet and Felipe Petroski Such and Dave Cummings and Matthias Plappert and Fotios Chantzis and Elizabeth Barnes and Ariel Herbert-Voss and William Hebgen Guss and Alex Nichol and Alex Paino and Nikolas Tezak and Jie Tang and Igor Babuschkin and Suchir Balaji and Shantanu Jain and William Saunders and Christopher Hesse and Andrew N. Carr and Jan Leike and Josh Achiam and Vedant Misra and Evan Morikawa and Alec Radford and Matthew Knight and Miles Brundage and Mira Murati and Katie Mayer and Peter Welinder and Bob McGrew and Dario Amodei and Sam McCandlish and Ilya Sutskever and Wojciech Zaremba},
  year={2021},
  eprint={2107.03374},
  archivePrefix={arXiv},
  primaryClass={cs.LG},
  doi={10.48550/arXiv.2107.03374},
}

@article{king1976symbolic,
author = {King, James C.},
title = {Symbolic execution and program testing},
year = {1976},
publisher = {Association for Computing Machinery},
volume = {19},
number = {7},
doi = {10.1145/360248.360252},
journal = {Commun. ACM},
month = jul,
pages = {385–394}
}

@inproceedings{clarke1976testing,
author = {Clarke, Lori A.},
title = {A program testing system},
year = {1976},
publisher = {Association for Computing Machinery},
doi = {10.1145/800191.805647},
booktitle = {Proceedings of the 1976 Annual Conference},
pages = {488–491}
}

@inproceedings{ui1,
  author={Andrica, Silviu and Candea, George},
  booktitle={2011 IEEE/IFIP 41st International Conference on Dependable Systems \& Networks (DSN)}, 
  title={WaRR: A tool for high-fidelity web application record and replay}, 
  year={2011},
  volume={},
  number={},
  pages={403-410},
  doi={10.1109/DSN.2011.5958253}
}

@inproceedings{ui2,
    author = {Biagiola, Matteo and Stocco, Andrea and Ricca, Filippo and Tonella, Paolo},
    title = {Diversity-based web test generation},
    year = {2019},
    isbn = {9781450355728},
    publisher = {Association for Computing Machinery},
    address = {New York, NY, USA},
    url = {https://doi.org/10.1145/3338906.3338970},
    doi = {10.1145/3338906.3338970},
    booktitle = {Proceedings of the 2019 27th ACM Joint Meeting on European Software Engineering Conference and Symposium on the Foundations of Software Engineering},
    pages = {142–153},
    numpages = {12},
    location = {Tallinn, Estonia},
    series = {ESEC/FSE 2019}
}

@inproceedings{ui3,
  author={Gross, Florian and Fraser, Gordon and Zeller, Andreas},
  booktitle={2012 34th International Conference on Software Engineering (ICSE)}, 
  title={EXSYST: Search-based GUI testing}, 
  year={2012},
  volume={},
  number={},
  pages={1423-1426},
  doi={10.1109/ICSE.2012.6227232}
}

@incollection{ui4,
    title = {Approaches and Tools for Automated End-to-End Web Testing},
    editor = {Atif Memon},
    series = {Advances in Computers},
    publisher = {Elsevier},
    volume = {101},
    pages = {193-237},
    year = {2016},
    issn = {0065-2458},
    doi = {10.1016/bs.adcom.2015.11.007},
    url = {https://www.sciencedirect.com/science/article/pii/S0065245815000686},
    author = {Maurizio Leotta and Diego Clerissi and Filippo Ricca and Paolo Tonella},
    booktitle = {Advances in Computers},
}

@inproceedings{ui5,
  author={Yandrapally, Rahulkrishna and Sinha, Saurabh and Tzoref-Brill, Rachel and Mesbah, Ali},
  booktitle={2023 IEEE/ACM 45th International Conference on Software Engineering (ICSE)}, 
  title={Carving UI Tests to Generate API Tests and API Specification}, 
  year={2023},
  volume={},
  number={},
  pages={1971-1982},
  doi={10.1109/ICSE48619.2023.00167}
}

@article{pizzorno2024coverup,
    author = {Altmayer Pizzorno, Juan and Berger, Emery D.},
    title = {CoverUp: Effective High Coverage Test Generation for Python},
    year = {2025},
    issue_date = {July 2025},
    publisher = {Association for Computing Machinery},
    address = {New York, NY, USA},
    volume = {2},
    number = {FSE},
    url = {https://doi.org/10.1145/3729398},
    doi = {10.1145/3729398},
    journal = {Proc. ACM Softw. Eng.},
    month = jun,
    articleno = {FSE128},
    numpages = {23}
}

@article{sen2005cute,
    author = {Sen, Koushik and Marinov, Darko and Agha, Gul},
    title = {CUTE: a concolic unit testing engine for C},
    year = {2005},
    issue_date = {September 2005},
    publisher = {Association for Computing Machinery},
    address = {New York, NY, USA},
    volume = {30},
    number = {5},
    issn = {0163-5948},
    url = {https://doi.org/10.1145/1095430.1081750},
    doi = {10.1145/1095430.1081750},
    journal = {SIGSOFT Softw. Eng. Notes},
    month = sep,
    pages = {263–272},
    numpages = {10}
}

@article{schafer2023empirical,
    author = {Sch\"{a}fer, Max and Nadi, Sarah and Eghbali, Aryaz and Tip, Frank},
    title = {An Empirical Evaluation of Using Large Language Models for Automated Unit Test Generation},
    year = {2024},
    issue_date = {Jan. 2024},
    publisher = {IEEE Press},
    volume = {50},
    number = {1},
    issn = {0098-5589},
    url = {https://doi.org/10.1109/TSE.2023.3334955},
    doi = {10.1109/TSE.2023.3334955},
    journal = {IEEE Trans. Softw. Eng.},
    month = jan,
    pages = {85–105},
    numpages = {21}
}

@article{tufano2020unit,
  title={Unit test case generation with transformers and focal context},
  author={Tufano, Michele and Drain, Dawn and Svyatkovskiy, Alexey and Deng, Shao Kun and Sundaresan, Neel},
  journal={arXiv preprint arXiv:2009.05617},
  year={2020},
  doi={10.48550/arXiv.2009.05617},
}

@article{bareiss2022code,
  title={Code generation tools (almost) for free? a study of few-shot, pre-trained language models on code},
  author={Barei{\ss}, Patrick and Souza, Beatriz and d'Amorim, Marcelo and Pradel, Michael},
  journal={arXiv preprint arXiv:2206.01335},
  year={2022},
  doi={10.48550/arXiv.2206.01335},
}

@inproceedings{ferreira2025acceptance,
  title={Acceptance Test Generation with Large Language Models: An Industrial Case Study}, 
  author={Ferreira, Margarida and Viegas, Luís and Faria, João Pascoal and Lima, Bruno},
  booktitle={2025 IEEE/ACM International Conference on Automation of Software Test (AST)}, 
  year={2025},
  volume={},
  number={},
  pages={1-11},
  doi={10.1109/AST66626.2025.00007}
}

@inproceedings{masuda2025generating,
  author={Masuda, Satoshi and Kouzawa, Satoshi and Sezai, Kyousuke and Suhara, Hidetoshi and Hiruta, Yasuaki and Kudou, Kunihiro},
  booktitle={2026 International Conference on Artificial Intelligence, Computer, Data Sciences and Applications (ACDSA)}, 
  title={Generating High-Level Test Cases from Requirements Using LLM: an Industry Study}, 
  year={2026},
  volume={},
  number={},
  pages={1-9},
  doi={10.1109/ACDSA67686.2026.11467785}
}

@inproceedings{
    wang2025maintaincoder,
    title={MaintainCoder: Maintainable Code Generation Under Dynamic Requirements},
    author={Zhengren Wang and Rui Ling and Chufan Wang and Yongan Yu and Sizhe Wang and Zhiyu li and Feiyu Xiong and Wentao Zhang},
    booktitle={The Thirty-ninth Annual Conference on Neural Information Processing Systems},
    year={2025},
    url={https://openreview.net/forum?id=TIbQixHEFD},
    doi={10.52202/085713-0521}
}

@article{zhang2025empowering,
    author = {Zhang, Sai and Xing, Zhenchang and Guo, Ronghui and Xu, Fangzhou and Chen, Lei and Zhang, Zhaoyuan and Zhang, Xiaowang and Feng, Zhiyong and Zhuang, Zhiqiang},
    title = {Empowering Agile-Based Generative Software Development through Human-AI Teamwork},
    year = {2025},
    issue_date = {July 2025},
    publisher = {Association for Computing Machinery},
    address = {New York, NY, USA},
    volume = {34},
    number = {6},
    issn = {1049-331X},
    url = {https://doi.org/10.1145/3702987},
    doi = {10.1145/3702987},
    journal = {ACM Trans. Softw. Eng. Methodol.},
    month = jul,
    articleno = {156},
    numpages = {46}
}

@article{zhang2025qwen3,
  title={Qwen3 Embedding: Advancing Text Embedding and Reranking Through Foundation Models},
  author={Zhang, Yanzhao and Li, Mingxin and Long, Dingkun and Zhang, Xin and Lin, Huan and Yang, Baosong and Xie, Pengjun and Yang, An and Liu, Dayiheng and Lin, Junyang and others},
  journal={arXiv preprint arXiv:2506.05176},
  year={2025},
  doi={10.48550/arXiv.2506.05176},
}

@inproceedings{
    hong2023metagpt,
    title={Meta{GPT}: Meta Programming for A Multi-Agent Collaborative Framework},
    author={Sirui Hong and Mingchen Zhuge and Jonathan Chen and Xiawu Zheng and Yuheng Cheng and Jinlin Wang and Ceyao Zhang and Zili Wang and Steven Ka Shing Yau and Zijuan Lin and Liyang Zhou and Chenyu Ran and Lingfeng Xiao and Chenglin Wu and J{\"u}rgen Schmidhuber},
    booktitle={The Twelfth International Conference on Learning Representations},
    year={2024},
    url={https://openreview.net/forum?id=VtmBAGCN7o}
}

@inproceedings{korraprolu2025requirements,
    author = {Korraprolu, Brahma Reddy and Pinninti, Pavitra and Reddy, Y. Raghu},
    title = {Test Case Generation for Requirements in Natural Language - An LLM Comparison Study},
    year = {2025},
    isbn = {9798400714245},
    publisher = {Association for Computing Machinery},
    address = {New York, NY, USA},
    url = {https://doi.org/10.1145/3717383.3717389},
    doi = {10.1145/3717383.3717389},
    booktitle = {Proceedings of the 18th Innovations in Software Engineering Conference},
    articleno = {9},
    numpages = {5},
    location = {
    },
    series = {ISEC '25}
}

@inproceedings{kitsios2025automated,
  title={Automated Generation of Issue-Reproducing Tests by Combining LLMs and Search-Based Testing}, 
  author={Kitsios, Konstantinos and Castelluccio, Marco and Bacchelli, Alberto},
  booktitle={2025 40th IEEE/ACM International Conference on Automated Software Engineering (ASE)}, 
  year={2025},
  volume={},
  number={},
  pages={1982-1994},
  doi={10.1109/ASE63991.2025.00165}}

@article{harman:2010:tse,
  author={Harman, Mark and McMinn, Phil},
  journal={IEEE Transactions on Software Engineering}, 
  title={A Theoretical and Empirical Study of Search-Based Testing: Local, Global, and Hybrid Search}, 
  year={2010},
  volume={36},
  number={2},
  pages={226-247},
  doi={10.1109/TSE.2009.71}
}

@article{lukasczyk:2023:emse,
  author    = {Stephan Lukasczyk and Florian Kroi{\ss} and Gordon Fraser},
  title     = {An empirical study of automated unit test generation for python},
  journal   = {Empirical Software Engineering},
  volume    = {28},
  number    = {2},
  year      = {2023},
  doi       = {10.1007/s10664-022-10248-w},
}

@article{fraser:2015:developerstudy,
  author = {Fraser, Gordon and Staats, Matt and McMinn, Phil and Arcuri, Andrea and Padberg, Frank},
  title = {Does Automated Unit Test Generation Really Help Software Testers? A Controlled Empirical Study},
  year = {2015},
  doi = {10.1145/2699688},
  journal = {ACM Trans. Softw. Eng. Methodol.},
  month = sep,
  articleno = {23}
}

@inproceedings{martin2021:blackandwhite,
  author={Martin-Lopez, Alberto and Arcuri, Andrea and Segura, Sergio and Ruiz-Cortés, Antonio},
  booktitle={2021 IEEE 32nd International Symposium on Software Reliability Engineering (ISSRE)}, 
  title={Black-Box and White-Box Test Case Generation for RESTful APIs: Enemies or Allies?}, 
  year={2021},
  volume={},
  number={},
  pages={231-241},
  doi={10.1109/ISSRE52982.2021.00034}
}

@misc{kim_llamaresttest_2025,
    title = {{LlamaRestTest}: {Effective} {REST} {API} {Testing} with {Small} {Language} {Models}},
    shorttitle = {{LlamaRestTest}},
    url = {http://arxiv.org/abs/2501.08598},
    doi = {10.48550/arXiv.2501.08598},
    language = {en},
    urldate = {2025-07-08},
    publisher = {arXiv},
    author = {Kim, Myeongsoo and Sinha, Saurabh and Orso, Alessandro},
    month = apr,
    year = {2025},
    note = {arXiv:2501.08598 [cs]},
}

@inproceedings{viglianisi_resttestgen_2020,
    address = {Porto, Portugal},
    title = {{RESTTESTGEN}: {Automated} {Black}-{Box} {Testing} of {RESTful} {APIs}},
    copyright = {https://ieeexplore.ieee.org/Xplorehelp/downloads/license-information/IEEE.html},
    shorttitle = {{RESTTESTGEN}},
    url = {https://ieeexplore.ieee.org/document/9159077/},
    doi = {10.1109/icst46399.2020.00024},
    language = {en},
    urldate = {2025-07-09},
    booktitle = {2020 {IEEE} 13th {International} {Conference} on {Software} {Testing}, {Validation} and {Verification} ({ICST})},
    publisher = {IEEE},
    author = {Viglianisi, Emanuele and Dallago, Michael and Ceccato, Mariano},
    month = oct,
    year = {2020},
    pages = {142--152},
}

@misc{zhang_logiagent_2025,
    title = {{LogiAgent}: {Automated} {Logical} {Testing} for {REST} {Systems} with {LLM}-{Based} {Multi}-{Agents}},
    shorttitle = {{LogiAgent}},
    url = {http://arxiv.org/abs/2503.15079},
    doi = {10.48550/arXiv.2503.15079},
    language = {en},
    urldate = {2025-07-09},
    publisher = {arXiv},
    author = {Zhang, Ke and Zhang, Chenxi and Wang, Chong and Zhang, Chi and Wu, YaChen and Xing, Zhenchang and Liu, Yang and Li, Qingshan and Peng, Xin},
    month = mar,
    year = {2025},
    note = {arXiv:2503.15079 [cs]},
}

@inproceedings{atlidakis_restler_2019,
    address = {Montreal, QC, Canada},
    title = {{RESTler}: {Stateful} {REST} {API} {Fuzzing}},
    copyright = {https://ieeexplore.ieee.org/Xplorehelp/downloads/license-information/IEEE.html},
    isbn = {978-1-7281-0869-8},
    shorttitle = {{RESTler}},
    url = {https://ieeexplore.ieee.org/document/8811961/},
    doi = {10.1109/ICSE.2019.00083},
    language = {en},
    urldate = {2025-07-08},
    booktitle = {2019 {IEEE}/{ACM} 41st {International} {Conference} on {Software} {Engineering} ({ICSE})},
    publisher = {IEEE},
    author = {Atlidakis, Vaggelis and Godefroid, Patrice and Polishchuk, Marina},
    month = may,
    year = {2019},
    pages = {748--758},
}

@inproceedings{chen_chatunitest_2024,
    author = {Chen, Yinghao and Hu, Zehao and Zhi, Chen and Han, Junxiao and Deng, Shuiguang and Yin, Jianwei},
    title = {ChatUniTest: A Framework for LLM-Based Test Generation},
    year = {2024},
    isbn = {9798400706585},
    publisher = {Association for Computing Machinery},
    address = {New York, NY, USA},
    url = {https://doi.org/10.1145/3663529.3663801},
    doi = {10.1145/3663529.3663801},
    booktitle = {Companion Proceedings of the 32nd ACM International Conference on the Foundations of Software Engineering},
    pages = {572–576},
    numpages = {5},
    location = {Porto de Galinhas, Brazil},
    series = {FSE 2024}
}

@misc{anthropic2025claude45,
  author       = {{Anthropic}},
  title        = {Claude Sonnet 4.5 System Card},
  year         = {2025},
  month        = oct,
  howpublished = {Anthropic System Card},
  url          = {https://www.anthropic.com/claude-sonnet-4-5-system-card},
}

@misc{jain_testforge_2025,
    title = {{TestForge}: {Feedback}-{Driven}, {Agentic} {Test} {Suite} {Generation}},
    shorttitle = {{TestForge}},
    url = {http://arxiv.org/abs/2503.14713},
    doi = {10.48550/arXiv.2503.14713},
    language = {en},
    urldate = {2025-07-08},
    publisher = {arXiv},
    author = {Jain, Kush and Goues, Claire Le},
    month = mar,
    year = {2025},
    note = {arXiv:2503.14713 [cs]},
}

@inproceedings{gou2024critic,
  title     = {{CRITIC}: Large Language Models Can Self-Correct with Tool-Interactive Critiquing},
  author    = {Gou, Zhibin and Shao, Zhihong and Gong, Yeyun and Shen, Yelong and Yang, Yujiu and Duan, Nan and Chen, Weizhu},
  booktitle = {The Twelfth International Conference on Learning Representations (ICLR)},
  year      = {2024},
  url       = {https://openreview.net/forum?id=Sx038qxjek}
}

@article{arcuri2019restful,
author = {Arcuri, Andrea},
title = {RESTful API Automated Test Case Generation with EvoMaster},
year = {2019},
issue_date = {January 2019},
publisher = {Association for Computing Machinery},
volume = {28},
number = {1},
doi = {10.1145/3293455},
journal = {ACM Transactions on Software Engineering and Methodology (TOSEM)},
month = {jan},
articleno = {3},
numpages = {37}
}

@inproceedings{kim2023reinforcement,
author = {Kim, Myeongsoo and Sinha, Saurabh and Orso, Alessandro},
title = {Adaptive REST API Testing with Reinforcement Learning},
year = {2023},
publisher = {IEEE Press},
doi = {10.1109/ASE56229.2023.00218},
booktitle = {Proceedings of the 38th IEEE/ACM International Conference on Automated Software Engineering},
pages = {446–458},
numpages = {13}
}

@inproceedings{kim2023enhancing,
author = {Kim, Myeongsoo and Corradini, Davide and Sinha, Saurabh and Orso, Alessandro and Pasqua, Michele and Tzoref-Brill, Rachel and Ceccato, Mariano},
title = {Enhancing REST API Testing with NLP Techniques},
year = {2023},
publisher = {Association for Computing Machinery},
doi = {10.1145/3597926.3598131},
booktitle = {Proceedings of the 32nd ACM SIGSOFT International Symposium on Software Testing and Analysis},
pages = {1232–1243},
numpages = {12}
}

@inproceedings{martin2022online,
author = {Martin-Lopez, Alberto and Segura, Sergio and Ruiz-Cort\'{e}s, Antonio},
title = {Online Testing of RESTful APIs: Promises and Challenges},
year = {2022},
publisher = {Association for Computing Machinery},
doi = {10.1145/3540250.3549144},
booktitle = {Proceedings of the 30th ACM Joint European Software Engineering Conference and Symposium on the Foundations of Software Engineering},
pages = {408–420},
numpages = {13}
}

@article{merging,
    author = {Yang, Enneng and Shen, Li and Guo, Guibing and Wang, Xingwei and Cao, Xiaochun and Zhang, Jie and Tao, Dacheng},
    title = {Model Merging in LLMs, MLLMs, and Beyond: Methods, Theories, Applications, and Opportunities},
    year = {2026},
    issue_date = {June 2026},
    publisher = {Association for Computing Machinery},
    address = {New York, NY, USA},
    volume = {58},
    number = {8},
    issn = {0360-0300},
    url = {https://doi.org/10.1145/3787849},
    doi = {10.1145/3787849},
    journal = {ACM Comput. Surv.},
    month = feb,
    articleno = {216},
    numpages = {41}
}

@ARTICLE{finetuning2,
  author={Zhang, Quanjun and Fang, Chunrong and Zheng, Yi and Qian, Ruixiang and Yu, Shengcheng and Zhao, Yuan and Zhou, Jianyi and Yang, Yun and Zheng, Tao and Chen, Zhenyu},
  journal={IEEE Transactions on Software Engineering}, 
  title={Improving Retrieval-Augmented Deep Assertion Generation via Joint Training}, 
  year={2025},
  volume={51},
  number={4},
  pages={1232-1247},
  doi={10.1109/TSE.2025.3545970}
}

@article{finetuning1,
    author = {Shang, Ye and Zhang, Quanjun and Fang, Chunrong and Gu, Siqi and Zhou, Jianyi and Chen, Zhenyu},
    title = {A Large-Scale Empirical Study on Fine-Tuning Large Language Models for Unit Testing},
    year = {2025},
    issue_date = {July 2025},
    publisher = {Association for Computing Machinery},
    address = {New York, NY, USA},
    volume = {2},
    number = {ISSTA},
    url = {https://doi.org/10.1145/3728951},
    doi = {10.1145/3728951},
    journal = {Proc. ACM Softw. Eng.},
    month = jun,
    articleno = {ISSTA074},
    numpages = {23}
}

@article{nashid2025issue2test,
  title={Issue2Test: Generating Reproducing Test Cases from Issue Reports},
  author={Nashid, Noor and Bouzenia, Islem and Pradel, Michael and Mesbah, Ali},
  journal={arXiv preprint arXiv:2503.16320},
  year={2025},
  doi={10.48550/arXiv.2503.16320},
}

@inproceedings{tan2012tcomment,
  author={Tan, Shin Hwei and Marinov, Darko and Tan, Lin and Leavens, Gary T.},
  booktitle={2012 IEEE Fifth International Conference on Software Testing, Verification and Validation}, 
  title={@tComment: Testing Javadoc Comments to Detect Comment-Code Inconsistencies}, 
  year={2012},
  volume={},
  number={},
  pages={260-269},
  doi={10.1109/ICST.2012.106}
}

@inproceedings{wen2019large,
  author={Wen, Fengcai and Nagy, Csaba and Bavota, Gabriele and Lanza, Michele},
  booktitle={2019 IEEE/ACM 27th International Conference on Program Comprehension (ICPC)}, 
  title={A Large-Scale Empirical Study on Code-Comment Inconsistencies}, 
  year={2019},
  volume={},
  number={},
  pages={53-64},
  doi={10.1109/ICPC.2019.00019}
}

@inproceedings{yang2024swe,
 author = {Yang, John and Jimenez, Carlos and Wettig, Alexander and Lieret, Kilian and Yao, Shunyu and Narasimhan, Karthik and Press, Ofir},
 booktitle = {Advances in Neural Information Processing Systems},
 doi = {10.52202/079017-1601},
 editor = {A. Globerson and L. Mackey and D. Belgrave and A. Fan and U. Paquet and J. Tomczak and C. Zhang},
 pages = {50528--50652},
 publisher = {Curran Associates, Inc.},
 title = {SWE-agent: Agent-Computer Interfaces Enable Automated Software Engineering},
 url = {https://proceedings.neurips.cc/paper_files/paper/2024/file/5a7c947568c1b1328ccc5230172e1e7c-Paper-Conference.pdf},
 volume = {37},
 year = {2024}
}

@inproceedings{zhang2024codeagent,
    title = "{C}ode{A}gent: Enhancing Code Generation with Tool-Integrated Agent Systems for Real-World Repo-level Coding Challenges",
    author = "Zhang, Kechi  and
      Li, Jia  and
      Li, Ge  and
      Shi, Xianjie  and
      Jin, Zhi",
    editor = "Ku, Lun-Wei  and
      Martins, Andre  and
      Srikumar, Vivek",
    booktitle = "Proceedings of the 62nd Annual Meeting of the Association for Computational Linguistics (Volume 1: Long Papers)",
    month = aug,
    year = "2024",
    address = "Bangkok, Thailand",
    publisher = "Association for Computational Linguistics",
    url = "https://aclanthology.org/2024.acl-long.737/",
    doi = "10.18653/v1/2024.acl-long.737",
    pages = "13643--13658"
}

@article{huang2023agentcoder,
  title={Agentcoder: Multi-agent-based code generation with iterative testing and optimisation},
  author={Huang, Dong and Zhang, Jie M and Luck, Michael and Bu, Qingwen and Qing, Yuhao and Cui, Heming},
  journal={arXiv preprint arXiv:2312.13010},
  year={2023},
  doi={10.48550/arXiv.2312.13010},
}

@article{merrill2026terminal,
  title={Terminal-Bench: Benchmarking Agents on Hard, Realistic Tasks in Command Line Interfaces},
  author={Merrill, Mike A and Shaw, Alexander G and Carlini, Nicholas and Li, Boxuan and Raj, Harsh and Bercovich, Ivan and Shi, Lin and Shin, Jeong Yeon and Walshe, Thomas and Buchanan, E Kelly and others},
  journal={arXiv preprint arXiv:2601.11868},
  year={2026},
  doi={10.48550/arXiv.2601.11868},
}

@InProceedings{ahmedotter,
  title = 	 {Otter: Generating Tests from Issues to Validate {SWE} Patches},
  author =       {Ahmed, Toufique and Ganhotra, Jatin and Pan, Rangeet and Shinnar, Avraham and Sinha, Saurabh and Hirzel, Martin},
  booktitle = 	 {Proceedings of the 42nd International Conference on Machine Learning},
  pages = 	 {752--771},
  year = 	 {2025},
  editor = 	 {Singh, Aarti and Fazel, Maryam and Hsu, Daniel and Lacoste-Julien, Simon and Berkenkamp, Felix and Maharaj, Tegan and Wagstaff, Kiri and Zhu, Jerry},
  volume = 	 {267},
  series = 	 {Proceedings of Machine Learning Research},
  month = 	 {13--19 Jul},
  publisher =    {PMLR},
  url = 	 {https://proceedings.mlr.press/v267/ahmed25b.html}
}

@inproceedings{fu2025automatically,
    author = {Fu, Lingyue and Zhang, Bolun and Guan, Hao and Zhu, Yaoming and Qiu, Lin and Liu, Weiwen and Cao, Xuezhi and Cai, Xunliang and Zhang, Weinan and Yu, Yong},
    title = {Automatically Benchmarking LLM Code Agents through Agent-driven Annotation and Evaluation},
    year = {2026},
    isbn = {9798400723179},
    publisher = {International Foundation for Autonomous Agents and Multiagent Systems},
    address = {Richland, SC},
    url = {https://doi.org/10.65109/HJFB4234},
    doi = {10.65109/HJFB4234},
    booktitle = {Proceedings of the 25th International Conference on Autonomous Agents and Multiagent Systems},
    pages = {874–882},
    numpages = {9},
    location = {Paphos, Cyprus},
    series = {AAMAS '26}
}

@inproceedings{kang2022large,
    author = {Kang, Sungmin and Yoon, Juyeon and Yoo, Shin},
    title = {Large Language Models are Few-Shot Testers: Exploring LLM-Based General Bug Reproduction},
    year = {2023},
    isbn = {9781665457019},
    publisher = {IEEE Press},
    url = {https://doi.org/10.1109/ICSE48619.2023.00194},
    doi = {10.1109/ICSE48619.2023.00194},
    booktitle = {Proceedings of the 45th International Conference on Software Engineering},
    pages = {2312–2323},
    numpages = {12},
    location = {Melbourne, Victoria, Australia},
    series = {ICSE '23}
}

@software{artifact,
  author = {Tyler Stennett and Rangeet Pan and Bridget McGinn and Alessandro Orso and Saurabh Sinha},
  title  = {{Sakura Artifact}},
  year   = {2026},
  doi    = {10.5281/zenodo.21710100},
  key    = {artifact}
}

@software{artifact_abstraction,
  author = {Tyler Stennett and Rangeet Pan and Bridget McGinn and Alessandro Orso and Saurabh Sinha},
  title  = {{Sakura Artifact: Abstraction Examples}},
  year   = {2026},
  url    = {https://github.com/aster-test-generation/sakura/tree/133b373b88b01db772c0fc9ff4c9115c319ff72a/outputs/abstraction_examples},
  key    = {artifact-abstraction}
}

@software{artifact_motivating,
  author = {Tyler Stennett and Rangeet Pan and Bridget McGinn and Alessandro Orso and Saurabh Sinha},
  title  = {{Sakura Artifact: Motivating Example}},
  year   = {2026},
  url    = {https://github.com/aster-test-generation/sakura/tree/40fd787e5253b22228aa7530abd69779f3a2c5d4/outputs/motivating_example},
  key    = {artifact-motivating}
}

@misc{jacoco,
  author       = {{EclEmma Team}},
  title        = {{JaCoCo}: Java Code Coverage Library},
  year         = {2025},
  howpublished = {\url{https://www.jacoco.org/jacoco/}},
  note         = {Version 0.8.13}
}

@misc{openrouter,
  author = {{OpenRouter}},
  title = {OpenRouter API},
  year = {2026},
  howpublished = {\url{https://openrouter.ai/}},
  note = {Unified API gateway for large language models}
}

@misc{google_cloud,
  author = {{Google LLC}},
  title = {Google Cloud Platform},
  year = {2026},
  howpublished = {\url{https://cloud.google.com/}},
  note = {Cloud computing infrastructure and managed services}
}

@misc{devstral,
  author       = {{Mistral AI}},
  title        = {Devstral Small 2},
  year         = {2025},
  howpublished = {\url{https://huggingface.co/mistralai/Devstral-Small-2-24B-Instruct-2512}}
}

@misc{qwen3,
  author       = {{Qwen Team}},
  title        = {Qwen3-Coder},
  year         = {2025},
  howpublished = {\url{https://huggingface.co/Qwen/Qwen3-Coder-480B-A35B-Instruct}}
}

@article{comanici2025gemini,
  title={Gemini 2.5: Pushing the frontier with advanced reasoning, multimodality, long context, and next generation agentic capabilities},
  author={Comanici, Gheorghe and Bieber, Eric and Schaekermann, Mike and Pasupat, Ice and Sachdeva, Noveen and Dhillon, Inderjit and Blistein, Marcel and Ram, Ori and Zhang, Dan and Rosen, Evan and others},
  journal={arXiv preprint arXiv:2507.06261},
  year={2025},
  doi={10.48550/arXiv.2507.06261}
}

@misc{apache_bcel,
  author = {{Apache Software Foundation}},
  title = {Apache Byte Code Engineering Library (BCEL)},
  year = {2004},
  howpublished = {\url{https://commons.apache.org/proper/commons-bcel/}},
  note = {Java bytecode analysis and manipulation library}
}

@misc{claude_code,
  author       = {{Anthropic}},
  title        = {Claude Code},
  year         = {2025},
  howpublished = {\url{https://docs.claude.com/en/docs/claude-code/overview}},
}

@misc{gemini_cli,
  author       = {{Google}},
  title        = {Gemini {CLI}},
  year         = {2025},
  howpublished = {\url{https://github.com/google-gemini/gemini-cli}},
}

@misc{bdd,
  author = {Dan North},
  title = {Introducing Behavior-Driven Development},
  year = {2006},
  howpublished = {\url{https://dannorth.net/introducing-bdd/}},
}

@misc{apache,
  title={Apache Commons},
  author={Apache},
  url={https://github.com/apache},
  year={2026}
}

@misc{ibm:wca4eja,
  title={{watsonx Code Assistant for Enterprise Java Applications}},
  author={IBM},
  url={https://www.ibm.com/products/watsonx-code-assistant-for-enterprise-java-applications},
  year={2024}
}

@article{hayes2007answering,
    author = {Andrew F. Hayes and Klaus Krippendorff},
    title = {Answering the Call for a Standard Reliability Measure for Coding Data},
    journal = {Communication Methods and Measures},
    volume = {1},
    number = {1},
    pages = {77--89},
    year = {2007},
    publisher = {Routledge},
    doi = {10.1080/19312450709336664},
}

@book{gwet2014handbook,
  title={Handbook of Inter-Rater Reliability, 4th Edition: The Definitive Guide to Measuring The Extent of Agreement Among Raters},
  author={Gwet, K.L.},
  isbn={9780970806284},
  url={https://books.google.com/books?id=fac9BQAAQBAJ},
  year={2014},
  publisher={Advanced Analytics, LLC}
}

@article{quarfoot2016robust,
    author = {David Quarfoot and Richard A. Levine},
    title = {How Robust Are Multirater Interrater Reliability Indices to Changes in Frequency Distribution?},
    journal = {The American Statistician},
    volume = {70},
    number = {4},
    pages = {373--384},
    year = {2016},
    publisher = {Taylor \& Francis},
    doi = {10.1080/00031305.2016.1141708},
}

\end{document}